\documentclass[lettersize,journal]{IEEEtran}
\usepackage{amsmath,amsfonts}
\usepackage{algorithmic}
\usepackage{algorithm}
\usepackage{array}
\usepackage[caption=false,font=normalsize,labelfont=sf,textfont=sf]{subfig}
\usepackage{textcomp}
\usepackage{stfloats}
\usepackage{url}
\usepackage{verbatim}
\usepackage{graphicx}
\usepackage{cite}
\usepackage{multirow}
\usepackage{setspace}
\usepackage{amsmath}
\usepackage{amsthm}
\usepackage{amssymb}
\newtheorem{thm}{\protect\theoremname}
\newtheorem{prop}[thm]{\protect\propositionname}
\newtheorem{cor}[thm]{\protect\corollaryname}
\usepackage{babel}
\providecommand{\corollaryname}{Corollary}
\providecommand{\propositionname}{Proposition}
\providecommand{\theoremname}{Theorem}
\usepackage{cuted}
\allowdisplaybreaks

\usepackage[margin=0.2cm,figurename=Fig.]{caption}
\begin{document}

\author{Jin~Xu,  Xinyuan~Wu, Qisheng~Huang, and Peng~Sun 
\thanks{
Jin Xu is with the School of Management, Huazhong University of Science
and Technology, Wuhan 430074, China.}
\thanks{
Xinyuan Wu is with the School of Data Science, the Chinese University of Hong
Kong, Shenzhen, Shenzhen 518172, China. }
\thanks{
Qisheng Huang is with the School of Mechanical Engineering and Automation,
Harbin Institute of Technology, Shenzhen, Shenzhen 518071, China. (Corresponding author: Qisheng Huang. Email:
huangqisheng@hit.edu.cn) }
\thanks{
Peng Sun is with the College of Computer Science and Electronic
Engineering, Hunan University, Changsha 410082, China.}
}

\title{How should the server sleep? -- Age-energy tradeoff in sleep-wake
server systems}
\maketitle

\begin{abstract}
The surging demand for fresh information from various Internet of Things (IoT) applications requires oceans of data sampled from the physical environment to be transmitted and processed timely, which would lead to huge energy consumption. We investigate the sleep-wake strategies of servers in communication systems to reduce energy consumption while guaranteeing timely delivery of fresh information to users. Specifically, we investigate a multi-source single-server queueing system and propose a novel sleep-wake strategy called the Conditional Sleep (CS) scheme. Our analysis reveals that the CS scheme outperforms the widely-used Hysteresis Time (HT) and Bernoulli Sleep (BS) schemes in terms of Age of Information (AoI), while retaining the same energy consumption rate and Peak Age of Information (PAoI). We find that increasing the sleep period length leads to a reduction in energy consumption and an increase in PAoI, but it does not always increase AoI. Moreover, we show that using PAoI as the information freshness metric in designing sleep-wake strategies would make the server sleep infinitely long due to the PAoI being determined by first-order statistics. We further numerically show that having the bufferless system can achieve a better PAoI-energy tradeoff than the infinite buffer system when having a large sampling rate.
\end{abstract}

\section{Introduction\label{sec:Introduction}}

\subsection{Motivations}

Information freshness has garnered wide attention from academia and
industry nowadays due to its influential role in communication theory
and decision science. Users in the Internet of Things (IoT) applications
usually need fresh information for decision-making. For instance,
in smart manufacturing systems, the process controller will need timely
information about the production process to identify defects and anomalies,
which is necessary for on-the-fly decision-making \cite{iquebal2020consistent}.
An autonomous vehicle needs fresh information about the velocity,
acceleration, and trajectory of other nearby vehicles to guarantee
driving safety \cite{chen2022age}. The freshness of received information
is crucial in these applications. 

The age of information process has been used to characterize the information
freshness of users in communication systems \cite{kaul2012real}.
The age of information at time $t$ is defined as $\Delta(t)=t-\tau(t)$,
where $\tau(t)$ is the time-stamp of the freshest packet received
by the user by time $t$. So $\Delta(t)$ is the time period that
elapsed since the generation time of the freshest packet received
by the user. In this paper, we focus on two information freshness
metrics based on the age of information process $\Delta(t)$, namely
the Averaged Age of Information (AoI) which characterizes the average
value of the age process, and the Peak Age of Information (PAoI) which
is the average value of the age peaks \cite{costa2016age}. Both AoI
and PAoI measure how timely the users are informed about the physical
process when receiving a sequence of data updates. A small AoI or
PAoI usually indicates that the user receives the information timely.

To guarantee that users have a small AoI or PAoI, the sampling devices
such as cameras and video monitors need to sample the physical environment
frequently. This would cause large traffic for communication systems
as the sampled information must be transmitted or processed before
useful information is extracted. The transmission and processing of
the sampled high-resolution images and videos would then bring substantial
energy consumption for the communication systems (e.g., access points
and base stations) and computing entities (e.g., processors and servers)
\cite{richter2009energy,niu2015characterizing}. Sleep-wake strategies
are thus needed to reduce energy consumption for base stations and
servers while guaranteeing information freshness for users.

Having the server sleeping could also result in system inefficiency
since the sleeping base stations or servers may not wake up as soon
as new transmission or processing requests occur. Most previous studies
focused on characterizing the delay-energy tradeoff of sleep-wake
strategies in different communication systems, i.e., how much energy
can be traded off by a tolerable transmission delay \cite{liu2015small,pei2017energy,niu2015characterizing,verma2019energy,wang2020base}.
These sleep-wake strategies usually let the server transit to a low-energy-consuming
sleep mode whenever the traffic load is light. For instance, the base
station can sleep when no information is transmitted in the communication
systems \cite{niu2015characterizing}. The computing entity in a smart
manufacturing system can be turned off temporarily when no images
or videos need to be processed. However, when considering information
freshness as the primary concern of the users, will the sleep-wake
strategies be different from those considering delay as the objective? 

\emph{Information freshness} and \emph{delay} are metrics evaluating
the Quality of Service (QoS) from different perspectives. The delay
metric measures each data packet's sojourn time in the system, and
the information freshness measures how timely the users are informed.
The systems and policies for reducing delay are designed to transmit
\emph{every} data packet as quickly as possible. In contrast, to guarantee
that users receive timely updates for online decision-making, the
system needs to deliver the \emph{freshest} packet as soon as possible
instead of every data packet. This difference makes it unknown whether
the sleep-wake strategies balancing the delay-energy tradeoff are
applicable to the scenarios where energy and information freshness
are of primary interest.

This paper aims to analyze the tradeoff between information freshness
and energy consumption. Specifically, we aim to answer the following
research questions:
\begin{itemize}
\item How to evaluate the tradeoff between energy and information freshness
under different sleep-wake strategies?
\item How to design the strategy to achieve the optimal energy and information
freshness tradeoff?
\item What are the differences in strategies when using AoI and PAoI as
information freshness metrics?
\end{itemize}
The questions above are challenging for the following two main reasons:
1) It is common that multiple data sources share the server or base
station, and the systems with multiple data sources are generally
difficult to analyze (see \cite{yates2019age,xu2019towards}); 2)
The processing time for data packets and setup time for the base station/server
can be generally distributed, so we cannot apply analytical methods
that rely on exponential assumptions, such as continuous-time Markov
Chain analysis \cite{maatouk2018age} and Stochastic Hybrid System
analysis \cite{yates2019age}. 

\subsection{Key Contributions}

The main contributions of our paper are summarized as follows.
\begin{itemize}
\item \emph{Performance evaluation:} We model the communication system where
the users are interested in fresh information as a bufferless queueing
system with multiple data sources. We use a renewal-type analysis
to provide the closed-form expressions for AoI, PAoI, and energy consumption.
The closed-form expressions can be used in understanding the age-delay
tradeoff and providing insights for communication protocol designs.
This analytical method can also be adopted in other communication
systems to evaluate the information freshness.
\item \emph{New Sleep-wake Strategy}: We propose a novel idling scheme called
Conditional Sleep (CS) scheme to improve information freshness. The
idea of CS is to reduce the variation of sleeping periods, which is
different from two schemes balancing the delay-energy tradeoff, namely
Hysteresis Time (HT) scheme and Bernoulli Sleep (BS) scheme. Compared
with HT and BS, the CS scheme achieves the same PAoI and energy consumption
but a significantly smaller AoI. We show that the idea of CS can be
extended to different communication systems to reduce AoI. 
\item \emph{Age-energy tradeoff}: We show that a longer sleeping period
leads to lower energy consumption for the server and higher PAoI for
users. Counter-intuitively, we find that AoI does not always increase
with the extension of the sleeping period. We provide conditions for
this particular phenomenon, and reveal that the result is due to the
definition of the AoI. We further show that the result is not restricted
to the bufferless system that we analyze. Our analytical results provide
practitioners with guidance to evaluate how much energy can be saved
by compromising users' information freshness.
\item \emph{Difference between metrics AoI and PAoI}: We find that minimizing
the energy consumption under a PAoI constraint would lead to an infinitely
long sleeping period at the optimum. The energy minimization problem
under the AoI constraint does not have this issue. Our analysis reveals
that the difference is due to PAoI being determined by first-order
statistics of idling and sleeping periods, and AoI being determined
by second-order statistics. 
\end{itemize}
We organize the rest of this paper as follows. In Section \ref{sec:Related-Work},
we introduce the work related to this study. In Section \ref{sec:System-Model},
we present the system model. We derive the closed-form expressions
using queueing analysis in Section \ref{sec:Queueing-Analysis}. In
Section \ref{sec:Optimal-Sleep-wake-Design}, we obtain the optimal
sleep-wake strategies based on the closed-form expressions for system
performance. Section \ref{sec:Numerical-Study} further conducts numerical
studies to develop insights. Finally, we provide concluding remarks
and discuss the future research in Section \ref{sec:Conclusion-and-Future}.

\section{Related Work\label{sec:Related-Work}}

We now review the literature related to the delay-energy tradeoff
in communications systems, sleep-wake strategies, information freshness
in queueing systems, and the age-energy tradeoff.

\subsection{Delay-energy Tradeoff in Communication Systems\label{subsec:Delay-energy-Tradeoff-in}}

Sleep-wake strategies have been designed and investigated in many
communication systems. For instance, Guo \emph{et al.} \cite{guo2016delay}
studied sleep-wake strategies in heterogeneous networks (HetNet) and
hyper-cellular networks (HCN). Pei \emph{et al.} \cite{pei2017energy}
investigated the sleep-wake base stations in ultra-dense networks
and modeled the system as an M/G/1/N processor sharing vacation queueing
system. Liu \emph{et al.} \cite{liu2015small} derived the coverage
probability, achievable rate, and energy efficiency for the sleep-wake
base stations in small cell networks. Other research studies like
\cite{verma2019energy,wu2015base,onireti2017analytical,feng2017base,feng2017boost}
also investigated sleep-wake strategies in different communication
networks. However, all these studies focused on the delay-energy tradeoff
without discussing the tradeoff between energy and information freshness. 

\subsection{Sleep-wake Strategies in Queueing Systems\label{subsec:Different-Sleep-wake-Strategies}}

Communication networks are usually modeled as queueing systems. A
sleep-wake strategy in queueing systems usually consists of two parts:
1) a \emph{wakeup scheme} that determines when the server should wake
up, and 2) an \emph{idling scheme} that determines when the server
should sleep. Different wakeup schemes in queueing systems have been
discussed in the literature, such as N-policy \cite{ke2003optimal},
single-sleep scheme \cite{liu2012equilibrium}, multiple-sleep scheme
\cite{onireti2017analytical}. Under the N-policy, the base station
continues sleeping until the queue accumulates $N$ data packets.
Under the single-sleep scheme, the server sleeps for a certain period
and then wakes up. For the multiple-sleep scheme, the server sleeps
for multiple periods until the system becomes non-empty. All these
studies \cite{ke2003optimal,liu2012equilibrium,onireti2017analytical}
assumed that the server would enter the sleeping period once the system
becomes empty.

Some studies also discussed the idling scheme in sleep-wake strategies.
Niu \emph{et al.} \cite{niu2015characterizing} and Guo \emph{et al.}
\cite{guo2016delay} discussed Hysteresis Time (HT) scheme under which
the server stays idling until either a threshold time is reached or
a new packet arrives. Studies like \cite{wu2013single,maraghi2009batch}
investigated the Bernoulli Sleep (BS) scheme, where the server takes
vacations with a probability after completing a task. However, these
studies mainly evaluated classic queueing performance metrics such
as mean delay, throughput, idling probability, and energy consumption
without considering information freshness.

\subsection{Information Freshness in Queueing Systems}

There are multiple studies that analyze the information freshness
in different queueing systems. Kaul \emph{et al.} \cite{kaul2012real}
investigated the AoI in M/M/1 system. Costa \emph{et al.} \cite{costa2016age}
further studied the AoI in single buffer systems, including the M/M/1/1,
M/M/1/2, and M/M/1/2{*} systems. They showed the advantage of dropping
redundant packets in minimizing AoI. Other recent studies considered
the information freshness in M/G/1/1 type systems with multiple data
streams, including \cite{najm2018status,dogan2021multi,moltafet2022moment,chen2022age}.
There are also papers investigating the information freshness in multi-hop
relay networks, including \cite{bedewy2019age,lou2020aoi,moradian2020age,champati2020statistical}.
The above studies mainly focus on the information freshness in different
queueing networks without considering the age-energy tradeoffs.

\subsection{Tradeoff between Energy and Information Freshness \label{subsec:Sleep-wake-Design-with}}

There are few recent studies that investigate the tradeoff between
energy and information freshness. Bedewy \emph{et al.} \cite{bedewy2020optimizing}
studied the sleep-wake scheduling for sensors to balance the information
freshness and energy tradeoff from the sampler's perspective. Huang
\emph{et al.} \cite{huang2021age} studied the information freshness
and energy tradeoff in fading channels, where strategies were designed
to minimize the weighted summation of AoI and energy consumption for
sensors. Several studies also considered energy consumption in information
sensing, including \cite{zhou2019joint,leng2019age,zheng2019age}.
However, most of the studies focused on energy consumption for sampling
devices without discussing the energy consumption for servers. Xu
and Chen \cite{xu2021Information} analyzed the PAoI in single-source
systems with the Last Come First Serve (LCFS) scheme and N-policy,
single-sleep, and multiple-sleep as sleep-wake strategies without
further discussing the information freshness and energy tradeoff.
It is still unclear how to design the server's sleep-wake strategy
to save energy while guaranteeing the information freshness for multiple
users.

In this work, we aim to design sleep-wake strategies for servers to
achieve optimal information freshness and energy tradeoff. The theoretical
and numerical results established in this work would provide useful
tools and insights for practitioners to better utilize communication
facilities.

\section{System Model\label{sec:System-Model}}

We will introduce the queueing model in Section \ref{subsec:Queueing-System}.
We will discuss how the server sleeps and wakes up in Section \ref{subsec:Idling-Schemes}
and Section \ref{subsec:Wakeup-Schemes}, respectively. We will then
provide information freshness and energy consumption metrics in Section
\ref{subsec:Information-Freshness-Metrics} and Section \ref{subsec:Energy-Consumption-Rate}.
Table \ref{tab:Notations} contains most of the notation used in this
paper. 

\begin{table}[t]
\begin{center}\footnotesize{%
\begin{tabular}{|c|>{\centering}p{7cm}|}
\hline 
Notation & Meaning\tabularnewline
\hline 
\hline 
$\lambda_{i}$ & Sampling rate of data source $i$\tabularnewline
\hline 
$\lambda$ & Total sampling rate $\lambda=\sum_{i=1}^{k}\lambda_{i}$\tabularnewline
\hline 
$H_{i}$ & Service time for packets from data source $i$\tabularnewline
\hline 
$U$ & Setup time\tabularnewline
\hline 
$D_{i}$ & Hysteresis time variable for HT after processing a source $i$ packet\tabularnewline
\hline 
$B_{i}$ & Threshold variable for CS when serving a source $i$ packet\tabularnewline
\hline 
$\theta_{i}^{Y}$ & Probability of sleeping after serving a source $i$ packet under idling
scheme $Y$, where $Y$ can be HT, BS, or CS\tabularnewline
\hline 
$N$ & The number of packets arriving during the sleeping period to wake
up the server under the N-policy\tabularnewline
\hline 
$W$ & The sleep period length for the single-sleep and multiple-sleep schemes\tabularnewline
\hline 
$P_{B}$ & Busy time energy consumption rate\tabularnewline
\hline 
$P_{ID}$ & Idling energy consumption rate\tabularnewline
\hline 
$P_{SL}$ & Sleeping energy consumption rate\tabularnewline
\hline 
$P_{ST}$ & Setup energy consumption rate\tabularnewline
\hline 
$P_{DT}$ & Energy consumption rate for detection under the multiple-sleep scheme\tabularnewline
\hline 
$F_{X}(u)$ & Cumulative distribution function (CDF) for random variable $X$\tabularnewline
\hline 
$X^{*}(s)$ & Laplace Stieltjes Transform (LST) of random variable $X$\tabularnewline
\hline 
$X^{*(k)}(s)$ & The $k^{th}$ derivative of $X^{*}(s)$\tabularnewline
\hline 
\end{tabular}}\end{center}
\caption{Notation\label{tab:Notations}}

\end{table}

\begin{figure}[t]
\centering{\includegraphics[scale=0.3]{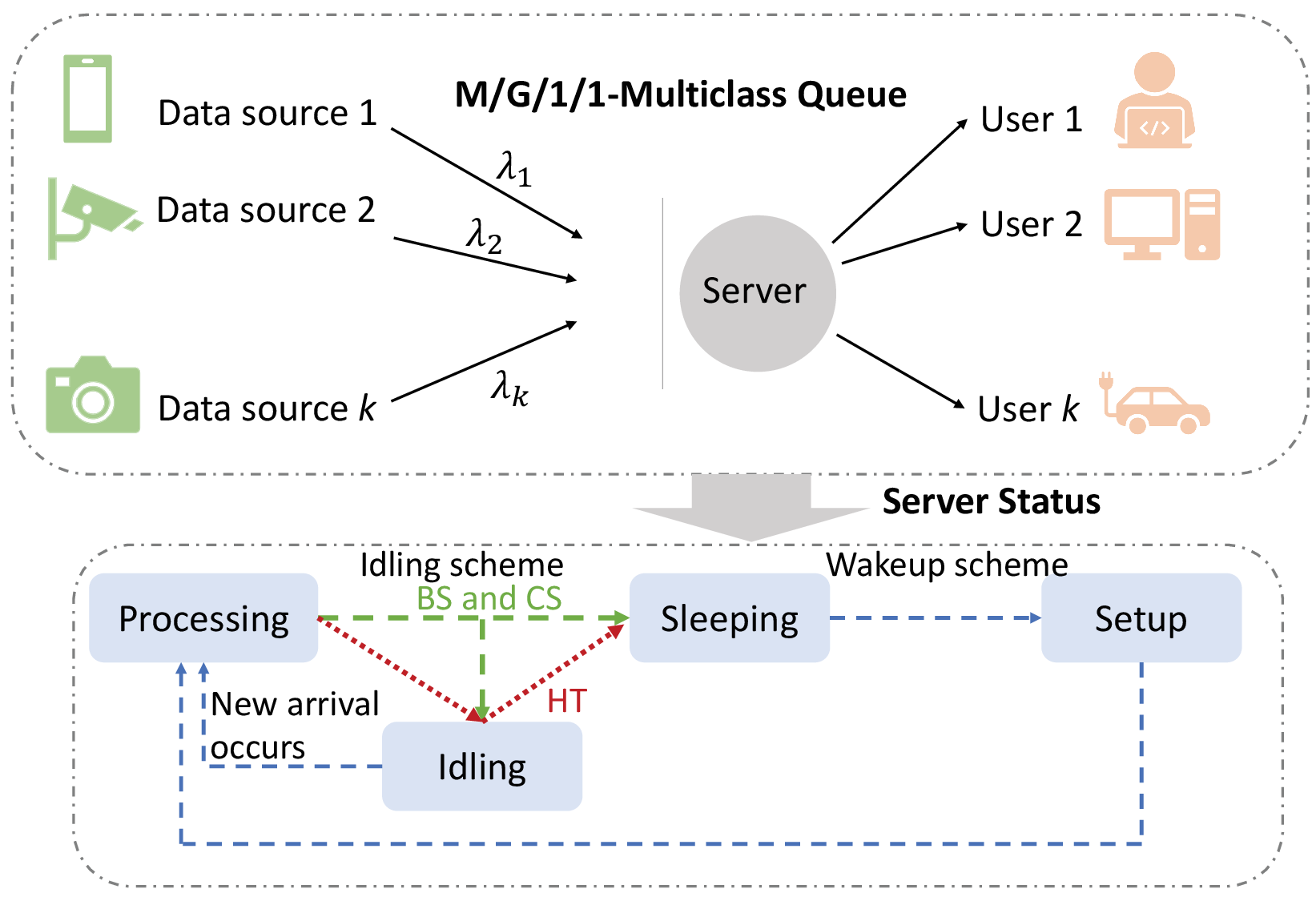}}

\caption{System model\label{fig:System-Model}}
\end{figure}

\subsection{Queueing System\label{subsec:Queueing-System}}

{We consider a system with $k$ data sources and a
single server (as shown in Fig. \ref{fig:System-Model}). Each data
source $i$ generates data packets independently following a Poisson
process with the sampling rate $\lambda_{i}$. Let $\lambda=\sum_{i=1}^{k}\lambda_{i}$
be the total sampling rate of all the data sources. The data packets
generated by each single data source have a particular destination
(user). Before being received by the user, each packet needs to be
processed by the server, and the server can process at most one packet
at a time. For packets from data source $i$, their processing times
$H_{i}$ are independent and identically distributed (i.i.d.). We
let $H_{i}$ be generally distributed. We consider a }{\emph{bufferless}}{{}
system that can hold at most one packet on the server. When the server
is processing, newly arrived packets will be rejected. The model we
study is thus an M/G/1/1 multi-class system with a sleep-wake server,
and it is an abstraction of multiple realistic communication systems,
as we explain in the following.}

{Our model can be an abstraction of a smart manufacturing
system with $k$ sampling devices, where each device samples a physical
process. The physical processes can correspond to the health of machine
tools, the condition of components, and the quality of workpieces
in machines. We assume the Poisson sampling process since it has been
widely used in practice and has multiple advantages in inferring the
statistical information about the physical process (see \cite{beutler1970alias,masry1978poisson,montanari2000random,karr2017point}).
Moreover, Poisson process can be an approximation of a variety of
sampling schemes and communication traffics (see \cite{kobayashi1977queueing,fuchs1969estimates}).
Assuming all the data are generated independently following Poisson
processes will also facilitate our analysis. }

{The sampled data packets could be high-resolution
photos and videos that need to be transmitted to the server (local
computing entity) and processed for extracting useful information,
such as detecting the defects of products and the degradation level
of the machine. The transmission and processing for each packet from
data source $i$ take $H_{i}$ amount of time. Once the processing
is completed, the system operator will immediately obtain the extracted
information and make online control decisions. We can regard the part
of the system operator that keeps tracking the status of process $i$
as user $i$.}

{There are several reasons why we consider the bufferless
system in our study. First, it has been proven to be more efficient
in improving information freshness than the infinite buffer system,
as redundant packets are dropped, enabling the delivery of fresh packets
in a timely manner (see \cite{moltafet2022moment,huang2015optimizing,najm2017status,chen2022age,costa2016age,inoue2019general,bedewy2019minimizing}).
This is particularly important for the smart manufacturing system,
where the system operator requires the freshest information for real-time
decision-making. Second, since the sampling devices usually need to
sample with high frequencies to improve information freshness, the
bufferless system is necessary to avoid queue congestion. With the
high sampling rates and long processing times, the infinite buffer
system may easily cause queue overflow. The bufferless system is experimented
with and testified to be effective in real-time control systems (see
\cite{chang2020age,kizilkaya2021age}). Third, we will demonstrate
in Section \ref{sec:Numerical-Study} that the bufferless system achieves
a better age-energy tradeoff than the infinite buffer system when
the sampling rate is large. Finally, most of the insights that we
develop for the bufferless system hold for the infinite buffer system,
as we will also show later.}

{The queueing system we study can also be the abstraction
of other communication systems where the information freshness is
of primary interest. For example, we can regard the server as the
base station in a wireless broadcast network. Data packets arrive
at the base station randomly, and they need to be transmitted by a
base station to the destination users (see \cite{kadota2019minimizing,chen2022age,hsu2019scheduling}).
In summary, we focus on the M/G/1/1 system for analysis since it is
an abstraction of multiple communication systems, and the results
and insights developed from this system can also be extended to other
systems. }

\subsection{Idling Schemes\label{subsec:Idling-Schemes}}

We consider the sleep-wake strategies consisting of two schemes: an
idling scheme that determines whether the server should sleep or idle
when no packet is waiting, and a wakeup scheme that determines when
the server should wake up after sleeping. In this paper, we consider
the following idling schemes:
\begin{itemize}
\item \emph{Hysteresis Time} (HT) scheme: The server would wait for $D_{i}$
amount of hysteresis time if a packet from source $i$ was just processed,
where $D_{i}$ is a random variable. If there is any arrival during
this period, the server will resume working immediately upon the packet's
arrival. If there is no arrival during this period $D_{i}$, the server
will sleep. This scheme was discussed in studies such as \cite{guo2016delay,xu2021Information}.
\item \emph{Bernoulli Sleep} (BS) scheme: After processing a packet from
source $i$, the server transits to the sleep mode with probability
$\theta_{i}^{BS}$, or stays idling with probability $1-\theta_{i}^{BS}$
until the next packet arrives. This scheme is easy to implement and
can be found in many queueing systems (see \cite{wu2013single,maraghi2009batch}).
\end{itemize}
The above HT and BS are idling schemes investigated in the literature.
Under HT, the server sleeps whenever the traffic is light (no arrival
within the idling period). The server under BS sleeps probabilistically
when the system becomes empty. The information freshness under these
two schemes has not been studied in the literature. Moreover, we propose
a new idling scheme called \emph{Conditional Sleep }(CS)\emph{ }scheme
as follows.
\begin{itemize}
\item \emph{Conditional Sleep} (CS) scheme: After processing a packet from
source $i$, if $H_{i}$ is the processing time of this packet and
$H_{i}<B_{i}$ with $B_{i}$ being a random variable, then the server
sleeps immediately. Otherwise, the server remains idling until the
next arrival occurs. For the convenience of analysis, we assume \textbf{$B_{i}$
}is exponentially distributed with rate $b_{i}$\footnote{It is also possible to select other distributions for $B_{i}$, but
the analysis will be more complicated. We leave the discussion on
the distribution of $B_{i}$ in Appendix \ref{sec:Discussion-for-the}
of the supplementary material. }. 
\end{itemize}
The motivation of CS is quite different from that of HT and BS. The
idea of CS is to avoid high age peaks in the future. When the server
has experienced a long processing time, the next age peak would be
high if the server further chooses to sleep. So that under CS, the
server only sleeps when its last processing time is small. We will
show that CS has the advantage in minimizing AoI over the other two
schemes.

\subsection{Wakeup Schemes\label{subsec:Wakeup-Schemes}}

We will mainly focus on the N-policy as the wakeup scheme in this
paper. Under the N-policy, the server will wake up whenever $N$ packets
have arrived during the current sleeping period \cite{guo2016delay,niu2015characterizing}.
We assume that the system only keeps the last packet among those $N$
packets. After waking up, the server would experience a setup time
$U$. During the setup time, newly arrived packets can still enter
the system, and the system keeps the last one. After the setup period,
the server starts processing.

Single-sleep and multi-sleep schemes are also widely used in practice
\cite{guo2016delay}. Under the single-sleep scheme, the server sleeps
for period $W$ and then sets up. If packets arrive during the sleeping
and setup periods, the server starts processing once the setup is
complete. Otherwise, the server will remain idling until the next
arrival. Under the multiple-sleep scheme, the server sleeps for multiple
periods with length $W$ until the system becomes non-empty. Then
the server experiences a setup period and starts processing. We also
assume for the single-sleep and multiple-sleep schemes that only the
freshest packet is kept in the system when the server is sleeping
or setting up. We will compare them with the N-policy
later.

\subsection{Information Freshness Metrics\label{subsec:Information-Freshness-Metrics}}

We now formally define the AoI and PAoI metrics. Let the \emph{age}
for user $i$ at time $t$ be $\Delta_{i}(t)=t-\tau_{i}(t)$, where
$\tau_{i}(t)$ is the generation time of the freshest packet received
by user $i$ before time $t$. By assuming the system is stationary
and ergodic, we define the \emph{AoI} of user $i$ (denoted as $\boldsymbol{E}[\Delta_{i}]$)
as the time-averaged value (stationary expectation) of $\Delta_{i}(t),$
i.e., $\boldsymbol{E}[\Delta_{i}]=\lim_{T\rightarrow\infty}\frac{1}{T}\int_{0}^{T}\Delta_{i}(t)dt$. 

The age process $\Delta_{i}(t)$ is piecewise linear, and each age
peak occurs right before a fresh packet is received by user $i$. By letting $A_{i,l}$ be the $l^{th}$ age peak of $\Delta_{i}(t)$
since time 0, we can define the \emph{PAoI} (denoted as $\boldsymbol{E}[A_{i}]$)
as the time-averaged (i.e., expected) age peaks, i.e., $\boldsymbol{E}[A_{i}]=\lim_{k\rightarrow\infty}\frac{1}{k}\sum_{l=1}^{k}A_{i,l}$.
Since we assume $0<\lambda_{i}<\infty$, as long as the server does
not sleep infinitely long, the AoI and PAoI are finite. A demonstrative graph of the age process is provided in Appendix
\ref{sec:Derivations-of-the}. Both AoI and PAoI can be used to characterize
information freshness as they share some similar properties \cite{xu2019towards,costa2016age}.
However, as we will show later, their properties in sleep-wake server
systems could be distinct.

\subsection{Energy Consumption Rate\label{subsec:Energy-Consumption-Rate}}

We aim to characterize the server's expected \emph{energy consumption
rate} $\boldsymbol{E}[P]$, i.e., the amount of energy consumed per
unit of time. Specifically, we assume that the server's energy consumption
rates are different over its statuses, namely $P_{B}$ for processing,
$P_{ID}$ for idling, $P_{SL}$ for sleeping, and $P_{ST}$ for setup.
The server's expected energy consumption rate $\boldsymbol{E}[P]$
is thus a function of $P_{B}$, $P_{IS}$, $P_{SL}$, and $P_{ST}$.
As we consider the scenario where sleeping can reduce the energy consumption
for the system, we assume that $P_{SL}<\min\{P_{B},P_{ID},P_{ST}\}$.
We also assume that $P_{ID}\leq P_{B}$, as the idling period usually
has a smaller energy consumption than the processing period (see \cite{guo2016delay,onireti2017analytical}).
For the multiple-sleep scheme, we assume there is an additional energy
consumption $P_{DT}$ to detect whether the system has a packet during
the sleeping period.

\section{Queueing Analysis\label{sec:Queueing-Analysis}}

This section focuses on the queueing analysis for the system under
different idling schemes. We will first introduce how the server sleeps
probabilistically under each idling scheme in Section \ref{subsec:Probabilistic-Sleep-under},
and then provide the closed-form expressions of AoI, PAoI, and energy
consumption rate in Section \ref{subsec:Closed-form-Expressions}.
Using the closed-form expressions, we further compare the performance
of the idling schemes in Section \ref{subsec:Idling-Scheme-Comparison}.

\subsection{Probabilistic Sleep under Idling Schemes\label{subsec:Probabilistic-Sleep-under}}

To derive the closed-form expressions of the system performance metrics,
we first introduce the concept of regenerative cycles. We define the
time span from processing a packet from data source $i$, to the time
when the server starts processing the next packet, as a Class $i$
regenerative cycle $V_{i}$. The idling schemes determine the probability
that the server sleeps within each regenerative cycle, and we now
characterize the sleeping probability for each idling scheme in the
following.

After processing a packet, the server under HT will remain idling
until either of the following two cases occurs: 1) an arrival occurs
before the hysteresis time $D_{i}$ is over, or 2) the server has
idled for time $D_{i}$. In the first case, $V_{i}$ will end at the
time the arrival occurs. In the second case, the server will further
experience a sleeping period and a setup period, and $V_{i}$ will
end when the setup period is over. Therefore, the probability that
the server sleeps within a regenerative cycle is $\theta_{i}^{HT}=\boldsymbol{P}(D_{i}\leq L)=D_{i}^{*}(\lambda)$,
where $L$ is the inter-arrival time of data packets. Under BS, each
$V_{i}$ begins with processing a packet. After processing the packet,
with probability $\theta_{i}^{BS}$ the server will sleep, and with
probability $1-\theta_{i}^{BS}$ a new $V_{j}$ will start. So the
probability that the server sleeps within a regeneration cycle is
$\theta_{i}^{BS}$. 

The server under CS only sleeps when the processing time $H_{i}$
is smaller than the threshold variable $B_{i}$. The idea of CS is
to remain idling when the server has processed a packet for a long
time, so that to reduce the peak age of the next regenerative cycle.
We assume \textbf{$B_{i}$ }is exponentially distributed with rate
$b_{i}$ for the convenience of analysis, so the sleeping probability
after serving a packet from source $i$ is $\theta_{i}^{CS}=\boldsymbol{P}(H_{i}<B_{i})=H_{i}^{*}(b_{i}).$

\subsection{Closed-form Expressions of the System Performance\label{subsec:Closed-form-Expressions}}

Based on the sleeping probability within each regenerative cycle,
we now provide the closed-form expressions of PAoI, AoI, and energy
consumption rate for these three idling strategies in the following
theorem. 
\begin{thm}
\label{thm:(1)-Under-HT,}When the N-policy is applied, the energy
consumption rate under HT, BS, or CS is 
\begin{align}
& \boldsymbol{E}[P^{Y}]=  \bigg\{\sum_{i=1}^{k}\frac{\lambda_{i}}{\lambda}\bigg[P_{B}\boldsymbol{E}[H_{i}]+\frac{1-\theta_{i}^{Y}}{\lambda}P_{ID}\nonumber\\
&+\theta_{i}^{Y}(\frac{N}{\lambda}P_{SL}+\boldsymbol{E}[U]P_{ST})\bigg]\bigg\}\nonumber \\
 & \bigg/\bigg\{\sum_{i=1}^{k}\frac{\lambda_{i}}{\lambda}\bigg[\boldsymbol{E}[H_{i}]+\theta_{i}^{Y}(\frac{N-1}{\lambda}+\boldsymbol{E}[U])+\frac{1}{\lambda}\bigg]\bigg\},\label{eq:0-1}
\end{align}
the PAoI for user $i$ is 
\begin{align}
\boldsymbol{E}[A_{i}^{Y}] & =\sum_{l=1}^{k}\frac{\lambda_{l}}{\lambda_{i}}\bigg[\boldsymbol{E}[H_{l}]+\theta_{l}^{Y}(\frac{N-1}{\lambda}+\boldsymbol{E}[U])\bigg]\nonumber \\
 & +\sum_{l=1}^{k}\theta_{l}^{Y}\frac{\lambda_{l}}{\lambda}\frac{1-U^{*}(\lambda)}{\lambda}+\frac{1}{\lambda_{i}}+\boldsymbol{E}[H_{i}],\label{eq:0-2}
\end{align}
where $Y$ can be HT, BS, and CS. The AoI under HT, BS, and CS is
given as
\begin{equation}
\boldsymbol{E}[\Delta_{i}^{Y}]=\frac{\boldsymbol{E}[I_{ii}^{2}]}{2\boldsymbol{E}[I_{ii}]}+\sum_{i=1}^{k}\theta_{i}^{Y}\frac{\lambda_{i}}{\lambda}\frac{1-U^{*}(\lambda)}{\lambda}+\boldsymbol{E}[H_{i}],\label{eq:0-3}
\end{equation}
where $I_{ii}^{*}(s)=\frac{V_{i}^{*}(s)\frac{\lambda_{i}}{\lambda}}{1-\sum_{l\neq i}\frac{\lambda_{l}}{\lambda}V_{l}^{*}(s)}$
and 
\begin{eqnarray}
V_{i}^{*}(s) =  \begin{cases}
\begin{aligned}
&  H_{i}^{*}(s)\bigg[\frac{\lambda}{s+\lambda}(1-D_{i}^{*}(s+\lambda))\\
 & +D_{i}^{*}(s+\lambda)(\frac{\lambda}{\lambda+s})^{N}U^{*}(s)\bigg], \end{aligned} &\mbox{under HT;}\\
\begin{aligned}&H_{i}^{*}(s)\bigg[\frac{\lambda}{s+\lambda}(1-\theta_{i}^{BS})\\
&+\theta_{i}^{BS}(\frac{\lambda}{\lambda+s})^{N}U^{*}(s)\bigg], \end{aligned}&\mbox{under BS;}\\
\begin{aligned}&\bigg[H_{i}^{*}(s)-H_{i}^{*}(s+b)\bigg]\frac{\lambda}{s+\lambda}\\
&+H_{i}^{*}(s+b)(\frac{\lambda}{\lambda+s})^{N}U^{*}(s),  \end{aligned} &\mbox{under CS}.
\end{cases}\label{eq:0-4}
\end{eqnarray}
\end{thm}
\begin{IEEEproof}
We leave the detailed proof to Appendix \ref{sec:Derivations-of-the}. 
\end{IEEEproof}
The closed-form expressions provided in Theorem \ref{thm:(1)-Under-HT,}
are useful for evaluating the theoretical performance of sleep-wake
strategies. Note that the impact of the idling schemes on energy consumption
rate and PAoI is only reflected through sleeping probabilities. When
$\theta_{i}^{CS}=\theta_{i}^{BS}=\theta_{i}^{HT}$ for $i=1,...,k$,
then HT, BS, and CS will result in the same energy consumption rate
and PAoI. The reason is that both energy consumption rate and PAoI
are first-order statistics of the idling period. In contrast, the
AoI under these idling schemes are different, as AoI depends on second-order
statistic $\boldsymbol{E}[I_{ii}^{2}]$.

We provide the performance of single-sleep and multiple-sleep schemes
in the following proposition, where we fix CS as the idling scheme.
The results for HT and BS as the idling scheme are also similar, and
we omit them here for conciseness.
\begin{prop}
\label{prop:When-fixing-CS}When fixing CS as the idling scheme, the
energy consumption rate under the single-sleep and multiple-sleep
schemes is 
\begin{align}
\boldsymbol{E}[P^{CS}]  = & \bigg\{\sum_{i=1}^{k}\frac{\lambda_{i}}{\lambda}\bigg[P_{B}\boldsymbol{E}[H_{i}]+\frac{1-\theta_{i}^{CS}}{\lambda}P_{ID}+\theta_{i}^{CS}P_{NP}\bigg]\bigg\}\nonumber \\
 &   \bigg/\bigg\{\sum_{i=1}^{k}\frac{\lambda_{i}}{\lambda}\bigg[\boldsymbol{E}[H_{i}]+\theta_{i}^{CS}T_{NP}+\frac{1-\theta_{i}^{CS}}{\lambda}\bigg]\bigg\},
\end{align}
with 
\begin{align}
P_{NP}  =  \begin{cases}
\begin{aligned}
&\boldsymbol{E}[W]P_{SL}+\boldsymbol{E}[U]P_{ST}\\
&+\frac{W^{*}(\lambda)U^{*}(\lambda)}{\lambda}P_{ID},  \mbox{ for single-sleep};\end{aligned}\\
\begin{aligned}&\frac{\boldsymbol{E}[W]P_{SL}}{1-W^{*}(\lambda)}+\frac{P_{DT}}{1-W^{*}(\lambda)}+\boldsymbol{E}[U]P_{ST},  \\
&\mbox{ for multiple-sleep}.\end{aligned}
\end{cases}
\end{align}
and 

\begin{align}
 T_{NP}  = \begin{cases}
\boldsymbol{E}[W]+\boldsymbol{E}[U]+\frac{W^{*}(\lambda)U^{*}(\lambda)}{\lambda}, & \mbox{for single-sleep};\\
\frac{\boldsymbol{E}[W]}{1-W^{*}(\lambda)}+\boldsymbol{E}[U], & \mbox{for multiple-sleep}.
\end{cases}
\end{align}
Under both the single-sleep and multiple-sleep schemes, the PAoI for
user $i$ is
\begin{eqnarray}
\boldsymbol{E}[A_{i}^{CS}] & = & \boldsymbol{E}[G_{i}]+\boldsymbol{E}[I_{ii}]+\boldsymbol{E}[H_{i}],\label{eq:3-1}
\end{eqnarray}
and the AoI for user $i$ is 
\begin{eqnarray}
\boldsymbol{E}[\Delta_{i}^{CS}] & = & \frac{\boldsymbol{E}[I_{ii}^{2}]}{2\boldsymbol{E}[I_{ii}]}+\boldsymbol{E}[G_{i}]+\boldsymbol{E}[H_{i}],\label{eq:4-1}
\end{eqnarray}
with $I_{ii}^{*}(s)=\frac{V_{i}^{*}(s)\frac{\lambda_{i}}{\lambda}}{1-\sum_{l\neq i}\frac{\lambda_{l}}{\lambda}V_{l}^{*}(s)}$, 
\begin{align}
&V_{i}^{*}(s)  =\nonumber\\
  & \begin{cases}
\begin{aligned} &\big[H_{i}^{*}(s)-H_{i}^{*}(s+b)\big]\frac{\lambda}{s+\lambda}+H_{i}^{*}(s+b)\big[W^{*}(s)U^{*}(s)\\
  &-\frac{s}{s+\lambda}W^{*}(s+\lambda)U^{*}(s+\lambda)\big], \mbox{ for single-sleep};\end{aligned}\\
\begin{aligned}&\big[H_{i}^{*}(s)-H_{i}^{*}(s+b)\big]\frac{\lambda}{s+\lambda}+\\
&H_{i}^{*}(s+b)\frac{W^{*}(s)-W^{*}(s+\lambda)}{1-W^{*}(s+\lambda)}U^{*}(s), \mbox{ for multiple-sleep}\end{aligned}.
\end{cases}
\end{align}
and
\begin{align}
\boldsymbol{E}[G_{i}]  =  \begin{cases}
\begin{aligned}
&\sum_{j=1}^{k}\theta_{j}^{CS}\frac{\lambda_{j}}{\lambda}\bigg[\frac{1-W^{*}(\lambda)U^{*}(\lambda)}{\lambda}+W^{*(1)}(\lambda)U^{*}(\lambda)\\
&+W^{*}(\lambda)U^{*(1)}(\lambda)\bigg],  \mbox{ for single-sleep};\end{aligned}\\
\begin{aligned}&\sum_{i=1}^{k}\theta_{i}^{CS}\frac{\lambda_{i}}{\lambda}\bigg[\frac{1}{\lambda}+U^{*}(\lambda)\frac{W^{*(1)}(\lambda)}{1-W^{*}(\lambda)}\bigg],\\
&  \mbox{ for multiple-sleep.}\end{aligned}
\end{cases}
\end{align}
\end{prop}
\begin{IEEEproof}
The derivation is based on the regenerative cycles, which is similar
to that for N-policy. We leave the detailed proof to Appendix \ref{sec:Performance-Metrics-for}
of the supplementary material.
\end{IEEEproof}
The closed-form expressions for the single-sleep and multiple-sleep
schemes turn out to be more complicated for those under the N-policy.
In the rest of the paper, we will mainly focus on the N-policy to
develop insights, and discuss the numerical results of the single-sleep
and multiple-sleep schemes in Section \ref{sec:Numerical-Study}.

\subsection{Idling Scheme Comparison\label{subsec:Idling-Scheme-Comparison}}

Since the multi-source scenario has a complicated closed-form expression
for each system performance metric, we now consider the single-source
scenario under N-policy to develop insights. The following theorem
shows the advantage of CS in minimizing AoI for the single-source
scenario.
\begin{thm}
\label{thm:If--is-1}For the single data source scenario with N-policy
as the wakeup scheme and $\theta_{1}^{CS}=\theta_{1}^{BS}=\theta_{1}^{HT}$
being fixed for HT, BS, and CS, then $\boldsymbol{E}[P^{CS}]=\boldsymbol{E}[P^{BS}]=\boldsymbol{E}[P^{HT}]$,
$\boldsymbol{E}[A_{1}^{CS}]=\boldsymbol{E}[A_{1}^{BS}]=\boldsymbol{E}[A_{1}^{HT}]$,
and $\boldsymbol{E}[\Delta_{1}^{CS}]\leq\boldsymbol{E}[\Delta_{1}^{BS}]\leq\boldsymbol{E}[\Delta_{1}^{HT}]$. 
\end{thm}
\begin{IEEEproof}
See Appendix \ref{sec:Proof-of-Theorem} for detailed proof.
\end{IEEEproof}
Theorem \ref{thm:If--is-1} proves that CS outperforms BS and HT in
minimizing AoI, and BS outperforms HT. An intuitive explanation for
Theorem \ref{thm:If--is-1} is as follows. AoI is determined by $\boldsymbol{E}[I_{ii}^{2}]$,
$\boldsymbol{E}[I_{ii}]$, $\boldsymbol{E}[G_{i}]$, and $\boldsymbol{E}[H_{i}]$.
When $\theta_{i}^{CS}=\theta_{i}^{BS}=\theta_{i}^{HT}$, CS, BS, and
HT have the same $\boldsymbol{E}[I_{ii}]$, $\boldsymbol{E}[G_{i}]$,
and $\boldsymbol{E}[H_{i}]$. The CS idling scheme has the smallest
$\boldsymbol{E}[I_{ii}^{2}]$, thus it results in the smallest AoI.

Under HT, the server would idle for a period after the system becomes
empty. If no arrival occurs during the idling period, then the server
falls asleep. In this case, the server would experience a large $I_{ii}$
that consists of a processing period, an idling period, a sleeping
period, and a setup period. If the server does not sleep, then $I_{ii}$
is short since it only consists of a processing period and an idling
period. The value of $\boldsymbol{E}[I_{ii}^{2}]$ is thus relatively
large as $I_{ii}$ is either too long or too short. Under BS, the
server would ``toss a coin'' to decide whether to sleep. If the decision
is to sleep, the server will sleep immediately without incurring an
idling period like HT. The second moment of $I_{ii}$ under BS is
thus smaller than that under HT. 

CS has an AoI smaller than BS and HT because it reduces the second
moment of $I_{ii}$. The server under CS will remain idling if the
packet processing time within this regenerative cycle turns out to
be large. It only sleeps when the processing time is short. This way
can prevent $I_{ii}$ from being either too large or too small, thus
reducing the second moment of $I_{ii}$. In Section \ref{sec:Numerical-Study},
we will use numerical study to show that CS has an advantage over
BS and HT in multi-source scenarios, and the advantage of CS could
be significant.

While Theorem \ref{thm:If--is-1} only focuses on the case when fixing
the N-policy as the wakeup scheme, we will show in Section \ref{sec:Numerical-Study}
that when the server adopts the single-sleep or multiple-sleep scheme,
CS achieves a smaller AoI than HT and BS while retaining the same
PAoI and energy consumption rate.\footnote{In fact, the advantage of CS over HT and BS also holds for M/G/1/LCFS
system. We leave the discussion to Appendix \ref{sec:Discussion-on-the}
of the supplementary material. }

\section{Optimal Parameter Selection for Sleep-wake Strategies \label{sec:Optimal-Sleep-wake-Design}}

In the previous section, we have characterized the performance of
different sleep-wake strategies. We now hope to answer a further question:
To achieve the optimal age-energy tradeoff, how often and how long
should the server sleep? Since CS has the advantage over BS and HT
in achieving a smaller AoI, we now fix CS as the idling scheme in
this section. We still focus on the N-policy as the wakeup scheme
due to its tractability. We aim to derive the threshold $b_{i}$ and
parameter $N$ to achieve the minimum energy consumption rate while
guaranteeing the information freshness for users. We will first provide
the conditions under which the AoI-energy tradeoff and PAoI-energy
tradeoff exist in Section \ref{subsec:Age-Energy-Tradeoff}, and then
discuss the optimal sleep-wake parameter selection in Section \ref{subsec:Optimization-Problem}. 

\subsection{Age-Energy Tradeoff\label{subsec:Age-Energy-Tradeoff}}

In this subsection, we will first introduce two useful corollaries
regarding the monotonicity of the PAoI and energy consumption rate,
and then discuss the difference between PAoI-energy tradeoff and AoI-energy
tradeoff. For notation simplicity, we omit the superscript and use
$\theta_{i}$ to denote the sleeping probability under CS.
\begin{cor}
\label{cor:-is-an}For a fixed $N$, $\boldsymbol{E}[A_{i}^{CS}]$
is an increasing function of $\theta_{i}$ for any $i$. For fixed
$\theta_{i}$ and $\min_{i\in\{1,...,k\}}\{\theta_{i}\}>0$, $\boldsymbol{E}[A_{i}^{CS}]$
is an increasing function of $N$. 
\end{cor}
\begin{IEEEproof}
Corollary \ref{cor:-is-an} holds because Equation (\ref{eq:0-2})
is a linear function of $\theta_{i}$ and $N$. 
\end{IEEEproof}
Corollary \ref{cor:-is-an} shows that enlarging the probability of
sleeping or extending the sleeping period length would increase the
PAoI. Note that Corollary \ref{cor:-is-an} holds for HT and BS as
well, as HT, BS, and CS have the same expressions for PAoI. In the
next corollary, we characterize how the energy consumption rate changes
as a function of $\theta_{i}$ and $N$.
\begin{cor}
\label{cor:When-,-then}When $P_{SL}<\min\{P_{ID},P_{ST},P_{B}\}$
and $\boldsymbol{\theta}=(\theta_{1},...,\theta_{k})$ is fixed with
$\min_{i\in\{1,...,k\}}\{\theta_{i}\}>0$, then $\boldsymbol{E}[P^{CS}]$
is decreasing on $N$. When $N$ is fixed, the minimal $\boldsymbol{E}[P^{CS}]$
is reached at either $\theta_{i}=0$ or $\theta_{i}=1$ for each $i$.
Furthermore, if $P_{ST}\leq P_{ID}$, then the minimal $\boldsymbol{E}[P^{CS}]$
is achieved at $\theta_{i}=1$ for all $i\in\{1,...,k\}$.
\end{cor}
\begin{IEEEproof}
See Appendix \ref{sec:Proof-of-Corollary} of the supplementary material
for detailed proof.
\end{IEEEproof}
Corollary \ref{cor:When-,-then} shows that increasing the sleeping
period length can always reduce energy consumption since the sleeping
mode has a lower energy consumption rate than the other server's statuses.
However, increasing the sleeping probabilities does not always reduce
energy consumption for a fixed $N$. The reason is that setup may
have a higher energy consumption than the sleeping period. Specifically,
when the sleeping period is short and the setup period is long, then
the energy saved during the sleeping period could be offset by the
energy consumed in the setup period. Sleeping frequently in this scenario
thus cannot reduce energy consumption. 

In the following corollary, we analyze the AoI in the single data
source scenario. We remove the subscript of the variables for notation
simplicity.
\begin{cor}
\label{cor:For-the-single}For the single-source scenario with $b=0$
in CS (i.e., $\theta=1$), if the coefficients of variation (CV) of
$H$ and $U$ are both smaller than 1 (i.e., $\frac{\sqrt{Var[H]}}{\boldsymbol{E}[H]}<1$
and $\frac{\sqrt{Var[U]}}{\boldsymbol{E}[U]}<1$), then AoI is increasing
with $N$. Otherwise, AoI may not always increase with $N$.
\end{cor}
\begin{IEEEproof}
See Appendix \ref{sec:Proof-of-Corollary-1} of the supplementary
material for detailed proof and discussion.
\end{IEEEproof}
Corollary \ref{cor:For-the-single} shows that lengthening the sleeping
period does not always enlarge AoI. This counter-intuitive result
is due to AoI being related to second-order statistics. As we show
in Theorem \ref{thm:(1)-Under-HT,}, AoI under CS is determined by
the term $\frac{\boldsymbol{E}[I_{ii}^{2}]}{2\boldsymbol{E}[I_{ii}]}$.
When extending the sleeping period, one can easily verify that both
$\boldsymbol{E}[I_{ii}]$ and $\boldsymbol{E}[I_{ii}^{2}]$ will increase.
However, their ratio does not always increase as $N$ increases. Moreover,
increasing the value of $b$ (the sleeping probability decreases in
this case) does not always increase the AoI. We will show that numerically
in Section \ref{subsec:Age-Energy-Tradeoff-1}.
\begin{table}[t!]
\centering{\footnotesize{%
\begin{tabular}{|>{\centering}p{1.5cm}|>{\centering}p{1.8cm}|>{\centering}p{1.8cm}|>{\centering}p{1.8cm}|}
\hline 
Actions & $\boldsymbol{E}[P^{CS}]$ & $\boldsymbol{E}[A^{CS}]$ & $\boldsymbol{E}[\Delta^{CS}]$\tabularnewline
\hline 
\hline 
Fix $\boldsymbol{b}$ and $b_{i}<\infty$, increase $N$ & Decrease & Increase & Non-monotone\tabularnewline
\hline 
Fix $N$, increase $b_{i}$ & Increase or decrease & Decrease & Non-monotone\tabularnewline
\hline 
\end{tabular}}}
\caption{Age-energy tradeoff under CS and N-policy\label{tab:Age-energy-tradeoff}}
\end{table}
Following Corollaries \ref{cor:-is-an}, \ref{cor:When-,-then}, and
\ref{cor:For-the-single}, we summarize the age-energy tradeoff in
Table \ref{tab:Age-energy-tradeoff}. When fixing the threshold vector
$\mathbf{b}=(b_{1},...,b_{k})$ finite and enlarging $N$, we find
that $\boldsymbol{E}[P^{CS}]$ decreases, $\boldsymbol{E}[A^{CS}]$
increases, but $\boldsymbol{E}[\Delta^{CS}]$ will not always increase
or decrease. When fixing the number $N$ and increasing the threshold
value $b_{i}$, the sleeping probability is reduced. In this case,
$\boldsymbol{E}[P^{CS}]$ either increases or decreases, depending
on the value of $P_{ST}$. The value of $\boldsymbol{E}[A^{CS}]$
decreases, but $\boldsymbol{E}[\Delta^{CS}]$ will not always increase
or decrease.

\subsection{Optimal Sleeping Probability and Sleeping Length\label{subsec:Optimization-Problem}}
We now investigate how to optimally
select the optimal sleeping probability and sleeping length to minimize
the energy consumption while keeping either PAoI or AoI bounded by
a threshold. 
\subsubsection{Optimization Problem Formulation}

We first formulate the optimization problem. Under CS, the energy
consumption rate $\boldsymbol{E}[P^{CS}(N,\mathbf{b})]$, PAoI $\boldsymbol{E}[A^{CS}(N,\mathbf{b})]$,
and AoI $\boldsymbol{E}[\Delta^{CS}(N,\mathbf{b})]$ are determined
by the parameter $N$ and vector $\mathbf{b}=(b_{1},...,b_{k})$.
When using PAoI as the information freshness metric, we have the following
problem:
\begin{eqnarray}
\mathbf{P1}: & \min & \boldsymbol{E}[P^{CS}(N,\mathbf{b})]\\
 & \mbox{s.t.} & \boldsymbol{E}[A_{i}^{CS}(N,\mathbf{b})]\leq\tau_{i};\\
 &  & b_{i}\geq0\mbox{ for }i\in\{1,...,k\};\\
 &  & N\in\mathcal{N}^{+}.
\end{eqnarray}
The parameter $\tau_{i}$ in \textbf{P1} is the PAoI requirement of
user $i$. So \textbf{P1 }is to minimize the energy consumption while
guaranteeing users' PAoI requirements. We can also define a problem
\textbf{P2} by substituting the PAoI constraint in \textbf{P1} with
a constraint enforcing the AoI requirement, i.e., $\boldsymbol{E}[\Delta_{i}^{CS}(N,\mathbf{b})]\leq\tau_{i}$,
with $\tau_{i}$ being the AoI requirement of user $i$. 

In order to characterize the difference in optimizing \textbf{P1 }and
\textbf{P2}, we begin by considering a single-source problem to develop
insights. Again, we remove the index of random variables in the single-source
scenario for notation simplicity. Since the impact of threshold $b$
reflects on the energy consumption rate and PAoI through the sleeping
probability $\theta$, we can rewrite the energy consumption rate
$\boldsymbol{E}[P^{CS}(N,\theta)]$ and PAoI $\boldsymbol{E}[A^{CS}(N,\theta)]$
as functions of the sleeping period length $N$ and sleeping probability
$\theta$. Since the integer variable $N$ in \textbf{P1 }makes the
problem difficult to analyze,\textbf{ }we relax the variable $N$
as a continuous variable\footnote{Letting $N$ be a real number also has practical implications. Suppose
$N\geq1$ is a real number, one can achieve $\boldsymbol{E}[A^{CS}(N,\theta)]$
and $\boldsymbol{E}[P^{CS}(N,\theta)]$ in the following way: Let
$\alpha=N-\lfloor N\rfloor$ such that the server with probability
$1-\alpha$ wakes up when accumulating $\lfloor N\rfloor$ packets,
and with probability $\alpha$ wakes up when accumulating $\lfloor N\rfloor+1$
packets. The optimal solution to \textbf{P1-relaxed }thus can be achieved
in practice.}, and the relaxed problem is denoted as problem \textbf{P1-relaxed}.
The optimal value of \textbf{P1-relaxed }is a lower bound of \textbf{P1},
and we will mainly focus on \textbf{P1-relaxed }to understand the
properties of \textbf{P1}.

\subsubsection{Optimal Solution Characterization \label{subsec:Optimal-Solution-Characterizatio}}

In the following theorem, we characterize the optimal solution to
\textbf{P1-relaxed}.
\begin{thm}
\label{thm:The-optimal-}The optimal $N$ and $\theta$ for the optimization
problem \textbf{P1-relaxed} belong to one of the three types: 
\end{thm}
\begin{itemize}
\item (Type 1) $\theta=\frac{\tau-\frac{1}{\lambda}-2\boldsymbol{E}[H]}{\frac{N}{\lambda}+\boldsymbol{E}[U]-\frac{U^{*}(\lambda)}{\lambda}}>0$
with $N\rightarrow\infty$; 
\item (Type 2) $\theta=1$ with $N=\lambda(\tau-\frac{1}{\lambda}-2\boldsymbol{E}[H]-\boldsymbol{E}[U]+\frac{U^{*}(\lambda)}{\lambda})$;
or 
\item (Type 3) $\theta=\frac{\tau-\frac{1}{\lambda}-2\boldsymbol{E}[H]}{\frac{1}{\lambda}+\boldsymbol{E}[U]-\frac{U^{*}(\lambda)}{\lambda}}$
with $N=1$.
\end{itemize}
\begin{IEEEproof}
See Appendix \ref{sec:Proof-for-Theorem} of the supplementary material
for detailed proof.
\end{IEEEproof}

Under the Type 1 solution in Theorem \ref{thm:The-optimal-}, the
server sleeps with a tiny but positive probability. Whenever the server
sleeps, it sleeps for infinitely long. Type 2 solution means that
the server sleeps deterministically, and the sleeping period is determined
by a number $N$ greater than 1. Under the Type 3 solution, the server
sleeps probabilistically and wakes up whenever a packet arrives during
the sleeping period.

When $N\rightarrow\infty$ and $\theta=\frac{\tau-\frac{1}{\lambda}-2\boldsymbol{E}[H]}{\frac{N}{\lambda}+\boldsymbol{E}[U]-\frac{U^{*}(\lambda)}{\lambda}}$,
both \textbf{P1 }and \textbf{P1-relaxed }have the same energy consumption.
Type 1 solution thus exists in \textbf{P1 }since \textbf{P1-relaxed
}is a lower bound of \textbf{P1}. Note that the Type 1 solution is
problematic in two aspects. First, the feasible region of \textbf{P1-relaxed
}is not a closed set. Type 1 solution is thus an asymptotic solution
located infinitely close to the boundary, but it cannot locate at
the boundary. It requires $N$ to be infinitely large and $\theta$
to be a positive number. Second, this solution is difficult to implement
in practice. Under this scheme, the server barely sleeps, but it sleeps
for an infinitely long period whenever it sleeps. No packet is processed
during the sleeping period, so the data receiver's age $\Delta(t)$
will become infinitely large.

\subsubsection{Occurrence of Different Types of Optimal Solutions\label{subsec:Occurrence-of-Different}}

\begin{figure}[t]
\center{
\subfloat[Optimal solutions for \textbf{P1-relaxed}]
{\includegraphics[scale=0.35]{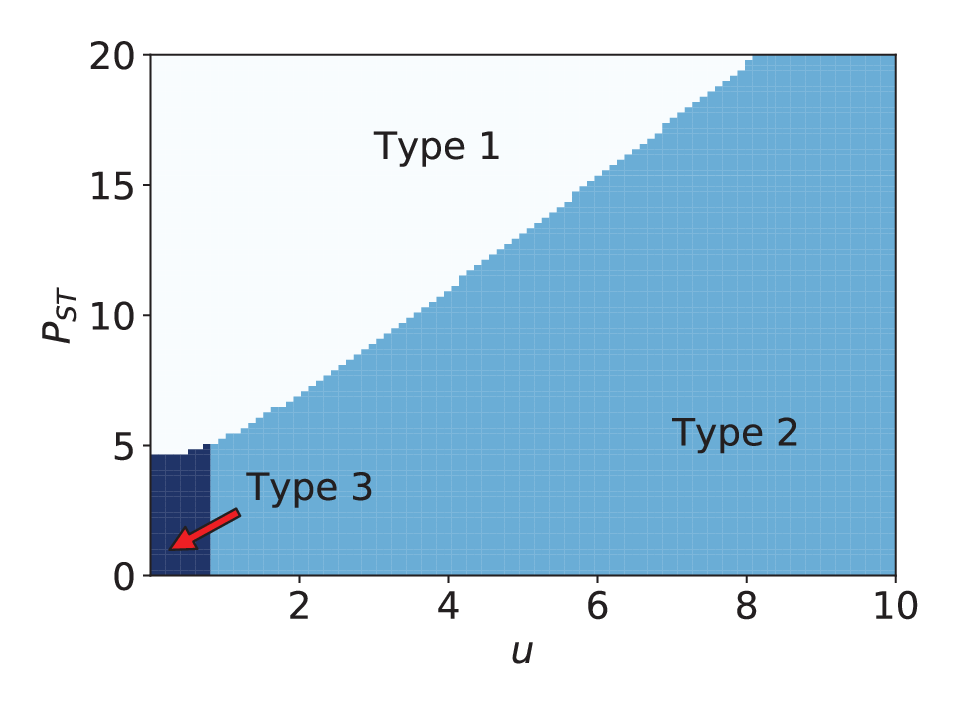}}

\subfloat[Optimal solutions for \textbf{P2-relaxed}]
{\includegraphics[scale=0.35]{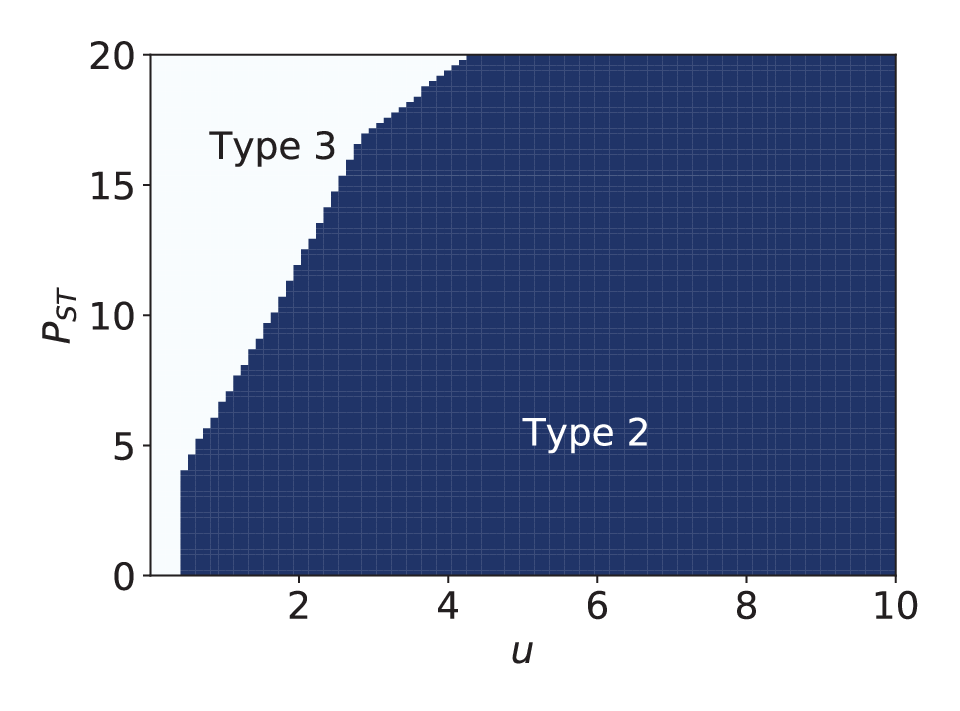}
}}
\caption{\label{fig:Different-Optimal-Solutions}Different types of optimal
solutions. $P_{B}=15$, $P_{ID}=7$, $P_{SL}=5$, $\lambda=0.9$.
$H\sim exp(1)$, $U\sim exp(u)$, $\tau=5$}
\end{figure}

We now discuss when different types of optimal solutions of \textbf{P1-relaxed}
would occur. Fig. \ref{fig:Different-Optimal-Solutions} provides
a numerical study to show when different types of optimal solutions
to \textbf{P1-relaxed} occur as the setup time $\boldsymbol{E}[U]=\frac{1}{u}$
and setup energy consumption $P_{ST}$ change. We can see from Fig.
\ref{fig:Different-Optimal-Solutions}(a) that Type 1 solution occurs
when $\boldsymbol{E}[U]$ and $P_{ST}$ are both large, as applying
Type 1 solution can effectively avoid setup. Moreover, we have the
following corollary for Type 1 solution. 

\begin{cor}
\label{cor:When-the-setup}When the setup time converges to zero,
i.e., $U\rightarrow0$, Type 1 solution in \textbf{P1-relaxed} will
not occur.
\end{cor}
\begin{IEEEproof}
See Appendix \ref{sec:Proof-of-Corollary-2} of the supplementary
material for detailed proof. 
\end{IEEEproof}
Fig. \ref{fig:Different-Optimal-Solutions}(a) also shows that when
$\boldsymbol{E}[U]$ and $P_{ST}$ are both small, we have the Type
2 optimal solution. Sleeping deterministically can guarantee the user's
PAoI requirement and also reduce energy consumption. When $\boldsymbol{E}[U]$
is large and $P_{ST}$ is small, we have the Type 3 optimal solution.
In this scenario, probabilistic sleeping can save energy and also
avoid creating a large $\boldsymbol{E}[A]$. 
When using AoI as the metric to measure information freshness, the
optimal $N$ in \textbf{P2} is always bounded. So \textbf{P2 }will
not have Type 1 solutions. One can easily verify from Equations (\ref{eq:0-3})
and (\ref{eq:0-4}) that if $N\rightarrow\infty$, then $\frac{\boldsymbol{E}[I_{11}^{2}]}{2\boldsymbol{E}[I_{11}]}\rightarrow\infty$.
Moreover, when relaxing the variable $N$ as a continuous variable,
we have Fig. \ref{fig:Different-Optimal-Solutions}(b) for \textbf{P2-relaxed},
where the optimal solution is either Type 2 (i.e., $\theta=1$)
or Type 3 (i.e., $\theta\in(0,1)$ and $N=1$). Specifically, we have
Type 2 solutions when both $u$ and $P_{ST}$ are large. Type 3 solutions
occur when $u$ is small but $P_{ST}$ is large. This result also
shows that although PAoI and AoI can both measure information
freshness, using them as information freshness metrics in optimization
problems would result in distinct solutions. The difference
comes from PAoI being determined by first-order statistics of idling
and sleeping periods, as shown in Equation (\ref{eq:0-2}). In contrast,
AoI is determined by the second moment of $I_{ii}$, as shown in Equation
(\ref{eq:0-3}).

\begin{figure}[t]
\center{\includegraphics[scale=0.35]{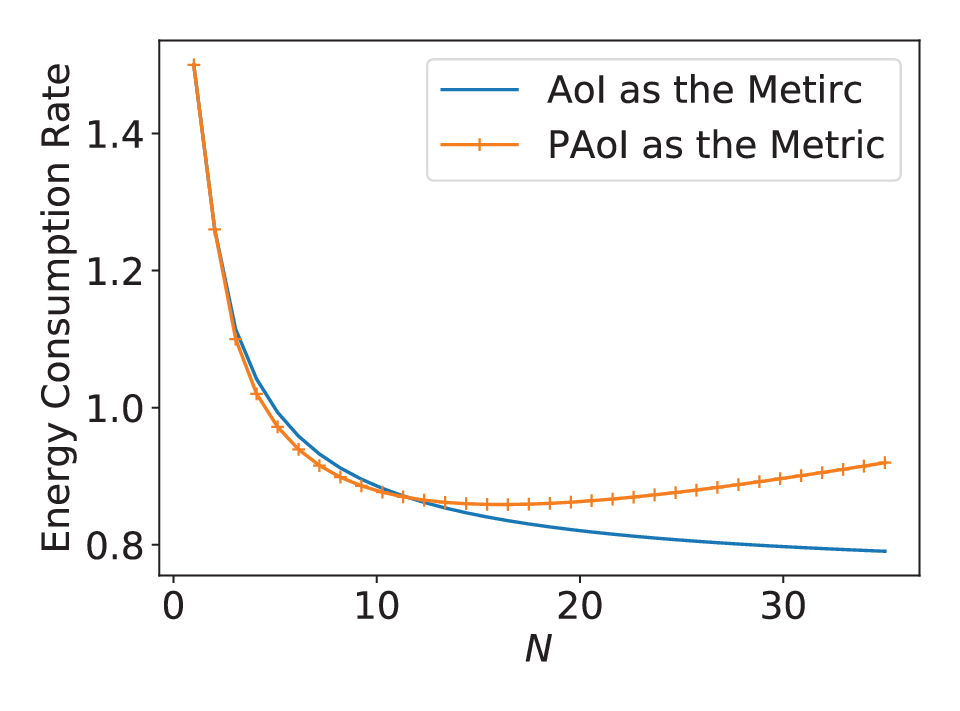}}
\caption{Optimal energy consumption rate. $\lambda_{i}=0.5$, $H_{i}\sim exp(1)$,
$V\sim exp(0.5)$ $\tau_{i}=30$, $P_{B}=2.1$, $P_{ID}=1.1$, $P_{SL}=0.3$,
$P_{ST}=1.8$. \label{fig:Optimal-Energy-Consumption-1}}
\end{figure}

As \textbf{P1} and \textbf{P2 }with multiple data sources are difficult
to be solved analytically, we present a numerical study for the case
of $k=5$ in Fig. \ref{fig:Optimal-Energy-Consumption-1}. For each
$N$ value in Fig. \ref{fig:Optimal-Energy-Consumption-1}, we solve
for the optimal $\boldsymbol{b}=(b_{i},i=1,...,5)$ to achieve the
minimal $\boldsymbol{E}[P^{CS}]$. We find that when using AoI as
the information freshness metric, the optimal $N$ is 15. However,
when using PAoI as the information freshness metric, the optimal $N$
could be larger than 35. As $P_{ST}$ is greater than $P_{ID}$ and
$P_{SL}$ in Fig. \ref{fig:Optimal-Energy-Consumption-1}, we can
infer that the optimal solution when using PAoI as the information
freshness metric (i.e., \textbf{P1}) is to avoid setup as much as
possible. That is, the optimal $\boldsymbol{\theta}$ could be very
tiny, and the optimal $N$ could be very large, similar to the Type
1 solution in \textbf{P1-relaxed}.

\section{Numerical Study\label{sec:Numerical-Study}}

In this section, we first use the numerical study to show the advantage
of CS in Section \ref{subsec:Idling-Strategy-Comparison}. We then
numerically discuss the AoI-energy and PAoI-energy tradeoffs in Section
\ref{subsec:Age-Energy-Tradeoff-1}. We will further compare our system
with LCFS service discipline in Section \ref{subsec:Comparison-between-LCFS}. 

\subsection{Idling Scheme Comparison\label{subsec:Idling-Strategy-Comparison}}

\begin{table*}
\footnotesize{\begin{center}%
\begin{tabular}{|>{\raggedright}p{2cm}|c|c|c|c|c|c|c|}
\hline 
\multicolumn{2}{|c|}{Schemes} & \multicolumn{2}{c|}{HT} & \multicolumn{2}{c|}{BS} & \multicolumn{2}{c|}{CS}\tabularnewline
\hline 
\hline 
\multicolumn{2}{|c|}{Metrics} & Simulation & Exact & Simulation & Exact & Simulation & Exact\tabularnewline
\hline 
\multirow{4}{2cm}{PAoI} & Source 1 & 59.4840 & \textbf{59.5945} & 59.5459 & \textbf{59.5945} & 59.6628 & \textbf{59.5944}\tabularnewline
\cline{2-8} \cline{3-8} \cline{4-8} \cline{5-8} \cline{6-8} \cline{7-8} \cline{8-8} 
 & Source 2 & 26.6623 & \textbf{26.6341} & 26.6663 & \textbf{26.6341} & 26.6279 & \textbf{26.6341}\tabularnewline
\cline{2-8} \cline{3-8} \cline{4-8} \cline{5-8} \cline{6-8} \cline{7-8} \cline{8-8} 
 & Source 3 & 18.7135 & \textbf{18.7056} & 18.7261 & \textbf{18.7056} & 18.7167 & \textbf{18.7065}\tabularnewline
\cline{2-8} \cline{3-8} \cline{4-8} \cline{5-8} \cline{6-8} \cline{7-8} \cline{8-8} 
 & Source 4 & 13.7891 & \textbf{13.7833} & 13.7880 & \textbf{13.7833} & 13.7898 & \textbf{13.7833}\tabularnewline
\hline 
\multirow{4}{2cm}{AoI} & Source 1 & 60.2203 & \textbf{60.4144} & 60.2598 & \textbf{60.3713} & 59.1983 & \textbf{59.3661}\tabularnewline
\cline{2-8} \cline{3-8} \cline{4-8} \cline{5-8} \cline{6-8} \cline{7-8} \cline{8-8} 
 & Source 2 & 27.8810 & \textbf{27.7874} & 27.7720 & \textbf{27.7443} & 26.8091 & \textbf{26.7391}\tabularnewline
\cline{2-8} \cline{3-8} \cline{4-8} \cline{5-8} \cline{6-8} \cline{7-8} \cline{8-8} 
 & Source 3 & 19.0326 & \textbf{19.0255} & 18.9734 & \textbf{18.9824} & 18.0155 & \textbf{17.9772}\tabularnewline
\cline{2-8} \cline{3-8} \cline{4-8} \cline{5-8} \cline{6-8} \cline{7-8} \cline{8-8} 
 & Source 4 & 15.4633 & \textbf{15.4366} & 15.4113 & \textbf{15.3935} & 14.4137 & \textbf{14.3884}\tabularnewline
\hline 
\multicolumn{2}{|c|}{Energy Consumption Rate} & 2.2191 & \textbf{2.2191} & 2.2192 & \textbf{2.2191} & 2.2191 & \textbf{2.2191}\tabularnewline
\hline 
\end{tabular}\end{center}}

\caption{\label{tab:Simulation-Results-v.s.}Simulation results v.s. exact
results. $\boldsymbol{\lambda}=(0.75,1.75,2.75,3.75)$ $H_{1}\protect\overset{d}{=}H_{3}\sim Gamma(0.3,\frac{1}{0.3^{2}}),$
$H_{2}\protect\overset{d}{=}H_{4}\sim Gamma(0.4,\frac{1}{0.4^{2}}),$
$\boldsymbol{\theta}=(0.3,0.4,0.4,0.3)$, $N=3$, $U\sim Gamma(5,1)$,
$P_{B}=2.1$, $P_{ID}=1.1$, $P_{SL}=0.3$, and $P_{ST}=2.5$. Simulation
is performed on a sample path of $3\times10^{6}$ regenerative cycles.}
\end{table*}

In Table \ref{tab:Simulation-Results-v.s.}, we compare the simulation
results with the analytical results that we provide in Theorem \ref{thm:(1)-Under-HT,}.
We find that the simulation results match the analytical ones. We
also observe that when the sleeping probabilities are equal, then
HT, BS, and CS result in the same energy consumption rate and PAoI,
while CS has the smallest AoI. This shows the advantage of using CS
as the idling scheme. 

\begin{figure*}[h]
\centering{\subfloat[N-policy with $N=1$]{\includegraphics[scale=0.35]{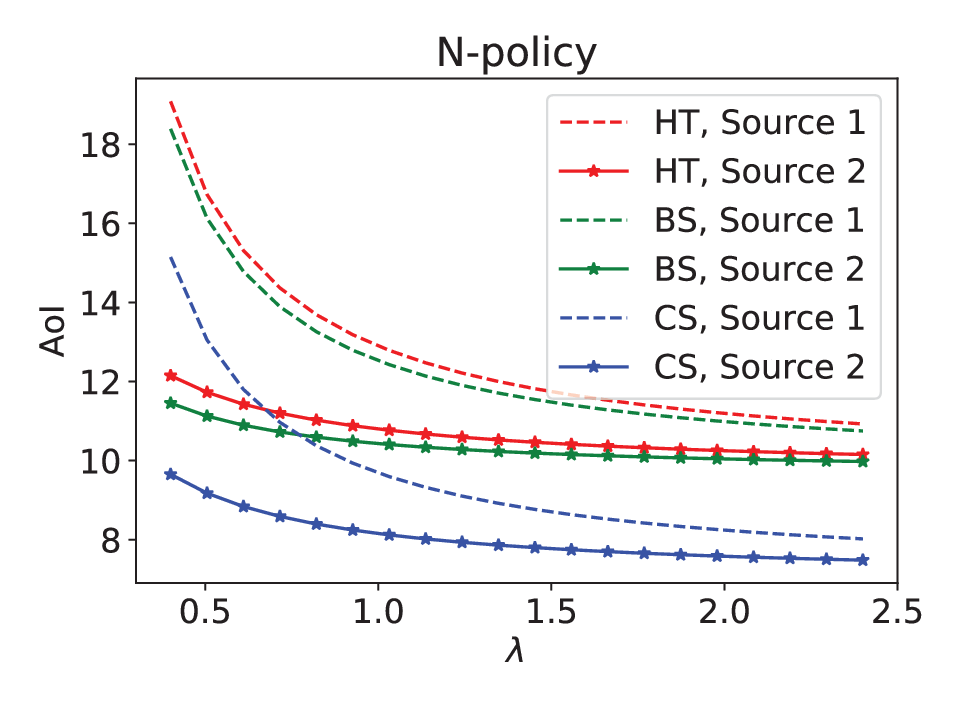}
}\subfloat[Single-sleep scheme with $W\sim exp(0.5)$]{\includegraphics[scale=0.35]{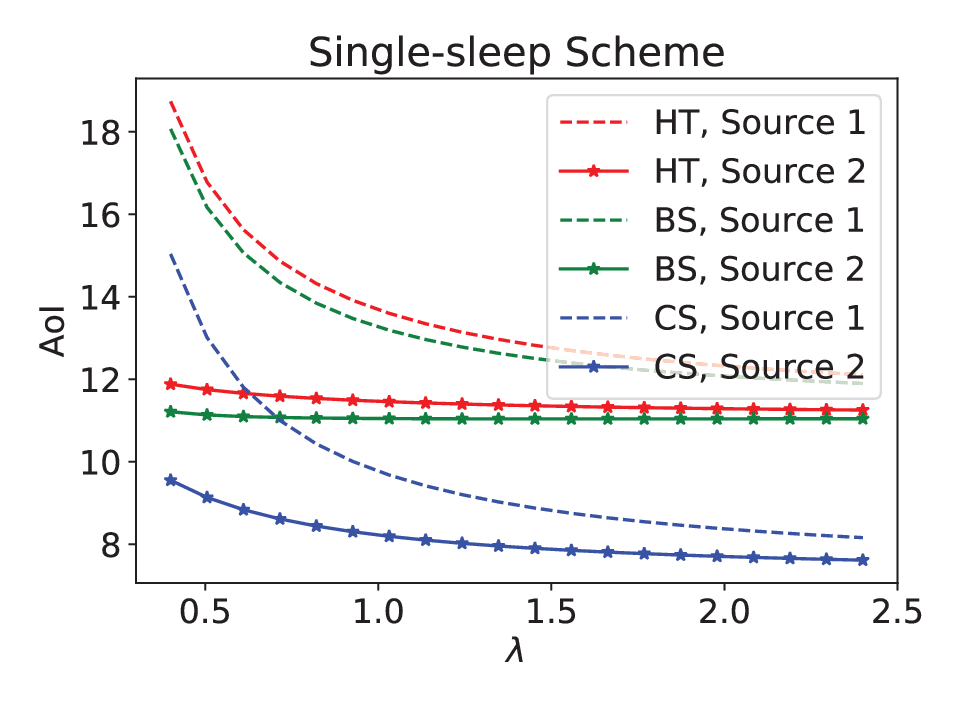}
}

\subfloat[Multiple-sleep scheme with $W\sim exp(0.5)$]{\includegraphics[scale=0.35]{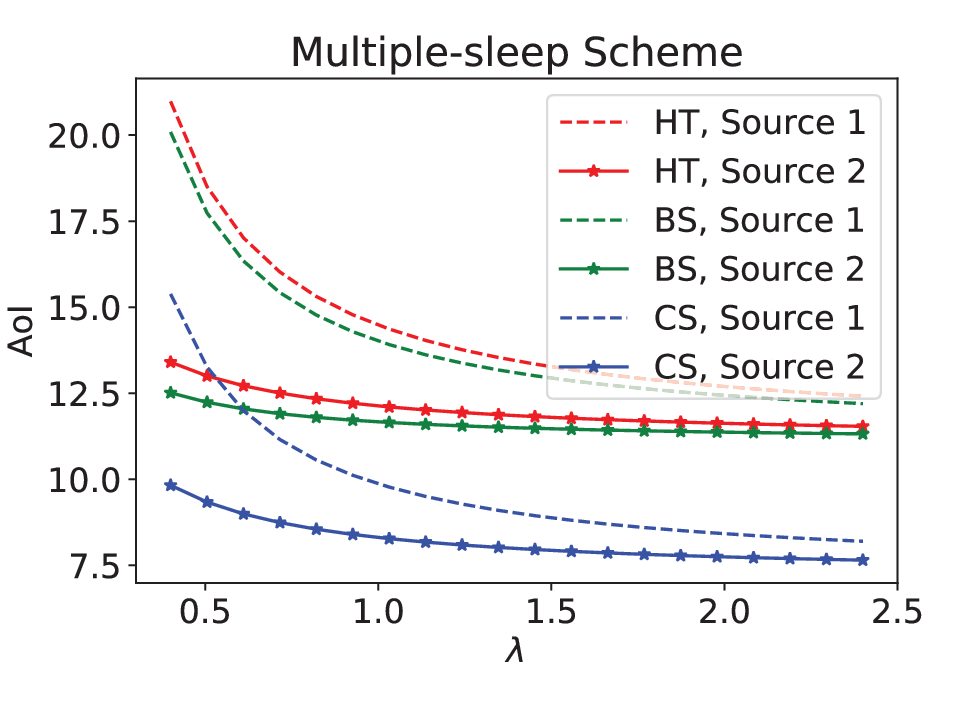}
}\subfloat[{Optimal $\boldsymbol{E}[P]$ in \textbf{P2 }under N-policy}]{\includegraphics[scale=0.35]{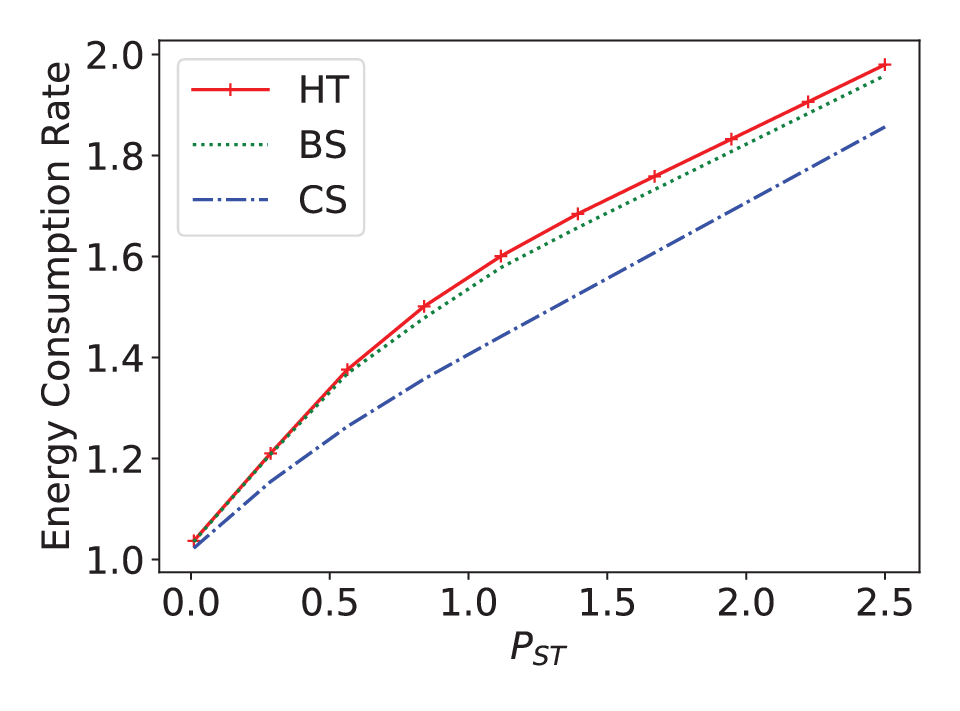}
}}
\caption{\label{fig:Idling-Strategy-Comparison}Idling scheme comparison with
$H_{1}\protect\overset{d}{=}H_{2}\sim Gamma(\frac{2}{5},\frac{25}{4})$
and $U\sim Gamma(10,1)$. For (a), (b), and (c), $\boldsymbol{\lambda}=(\lambda_{1},\lambda_{1}+0.1)$
and $\theta_{1}=\theta_{2}=0.3$. For (d), $\mathbf{\lambda}=(1.0,1.0)$,
$\tau=(16,16)$, $P_{B}=3.2,$ $P_{ID}=1.1,$ $P_{SL}=0.1$. }
\end{figure*}

In Fig. \ref{fig:Idling-Strategy-Comparison}, we compare the AoI
under different sleep-wake strategies for the multiple-source scenario.
In Fig. \ref{fig:Idling-Strategy-Comparison}, we let $D_{i}$ be
a constant for all $i=1,...,k$ in HT, so that $D_{i}=-\frac{\ln\theta_{i}}{\lambda}.$
For CS, we let $H_{i}$ be exponentially distributed with $b_{i}=\frac{1-\theta_{i}}{\theta_{i}\boldsymbol{E}[H_{i}]}$.
In Fig. \ref{fig:Idling-Strategy-Comparison}(a), we compare HT, BS,
and CS by fixing the wakeup scheme as the N-policy. In Fig. \ref{fig:Idling-Strategy-Comparison}(b)
and (c), we compare the idling schemes by fixing the wakeup scheme
as the single-sleep and multiple-sleep schemes. In Fig. \ref{fig:Idling-Strategy-Comparison}(d),
we compare the optimal energy consumption rates of \textbf{P2 }under
N-policy with different idling schemes.

We have several observations from Fig. \ref{fig:Idling-Strategy-Comparison}.
First, for the multiple-source scenario, CS still has a smaller AoI
than HT and BS for each data source, and the advantage of CS is significant.
The AoI for each data source under CS could be 30\% less than that
under HT. Second, the advantage CS has over HT and BS does not rely
on the wakeup scheme. When we replace the wakeup scheme with the single-sleep
scheme or multiple-sleep scheme, the advantage of CS is still significant,
as shown in Fig. \ref{fig:Idling-Strategy-Comparison}(b) and (c).
Third, when considering the energy minimization problem \textbf{P2}
with AoI as the constraint, the energy reduction by applying CS is
significant. As shown in Fig. \ref{fig:Idling-Strategy-Comparison}(d),
the energy consumption of CS could be 10\% less than that of HT and
BS. We can find from Fig. \ref{fig:Idling-Strategy-Comparison}(d)
that by applying CS, the energy consumption rate ranges from 1 to
1.8 for different $P_{ST}$ values. Compared with the energy consumption
rate when the server does not sleep (2.9571 in this case), the energy
reduction by applying the sleep-wake strategy could be 35\%-65\%.
This shows that sleep-wake strategies can lead to a substantial energy
reduction. 

\subsection{Tradeoff between Information Freshness and Energy Consumption \label{subsec:Age-Energy-Tradeoff-1}}

\begin{figure*}[h]
\subfloat[Energy consumption rate]{\includegraphics[scale=0.35]{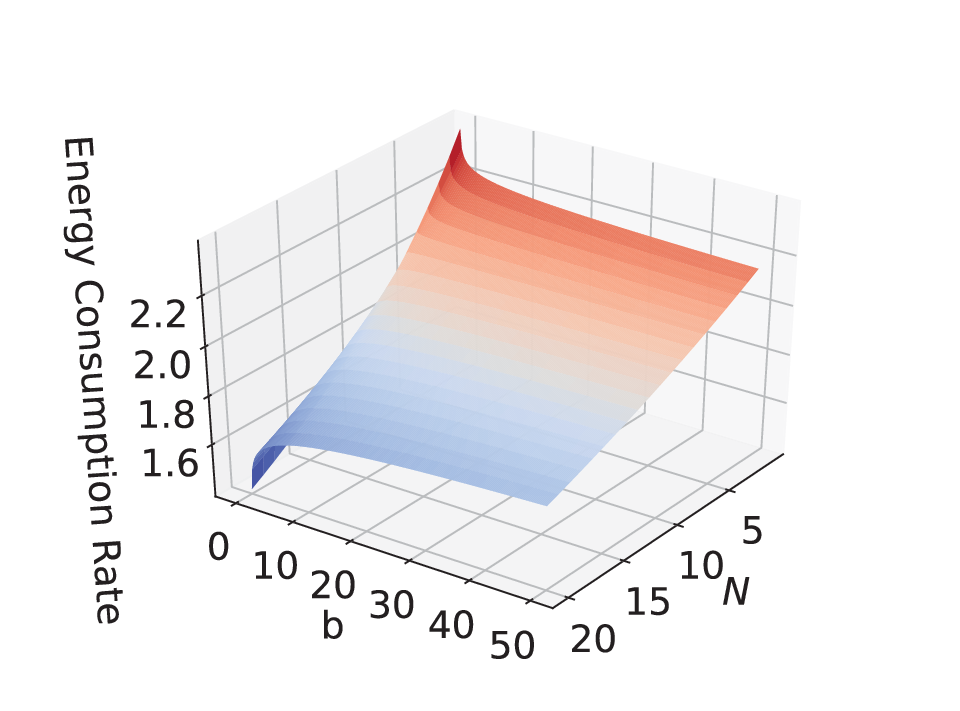}

}\subfloat[PAoI]{\includegraphics[scale=0.35]{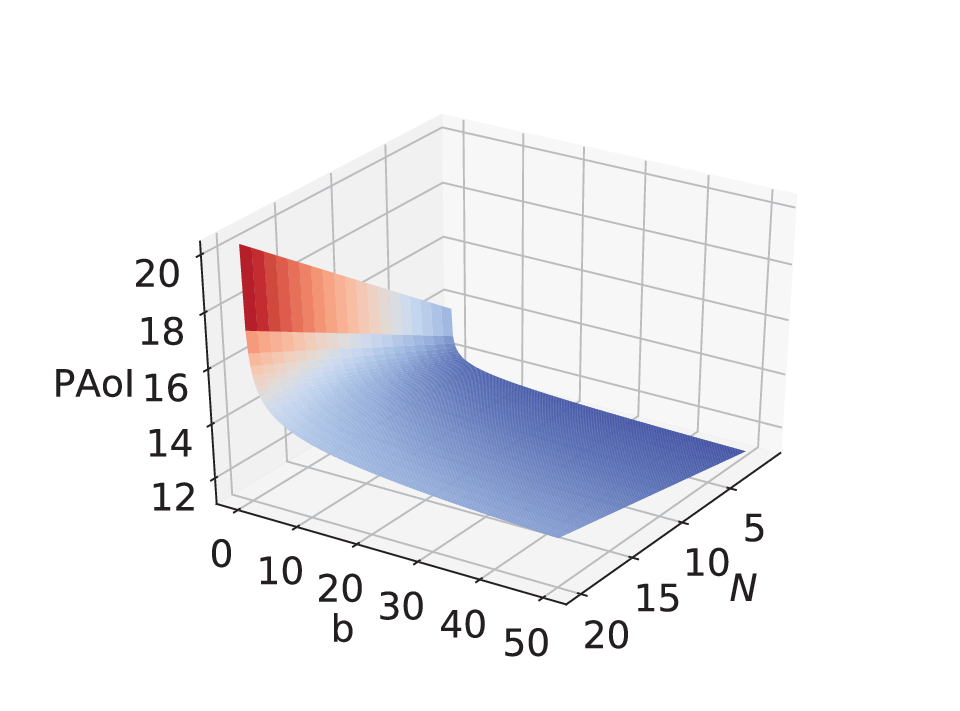}

}\subfloat[AoI]{\includegraphics[scale=0.35]{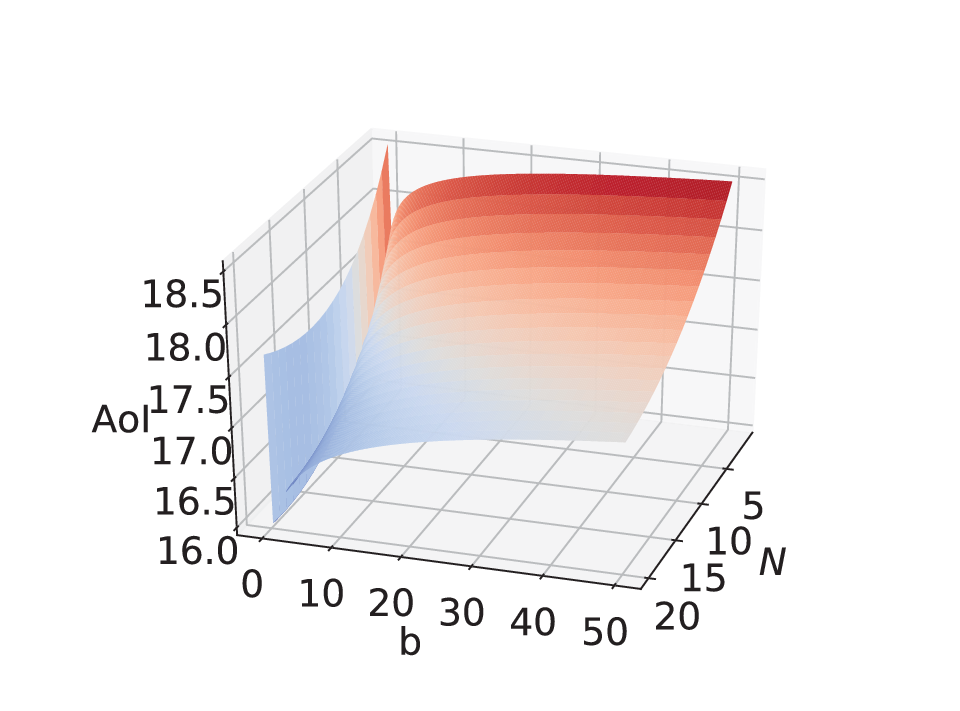}

}

\caption{Tradeoff between information freshness and energy. $\lambda=2.8,$
$H\sim Gamma(\frac{1}{5},25),$ $U\sim Gamma(0.3,\frac{1}{0.09})$,
$P_{B}=2.1,$ $P_{ID}=1.1,$ $P_{SL}=0.1,$ $P_{ST}=3.1$. \label{fig:Age-energy-Tradeoff}}
\end{figure*}

\begin{figure*}[h]
\subfloat[N-policy]{\includegraphics[scale=0.35]{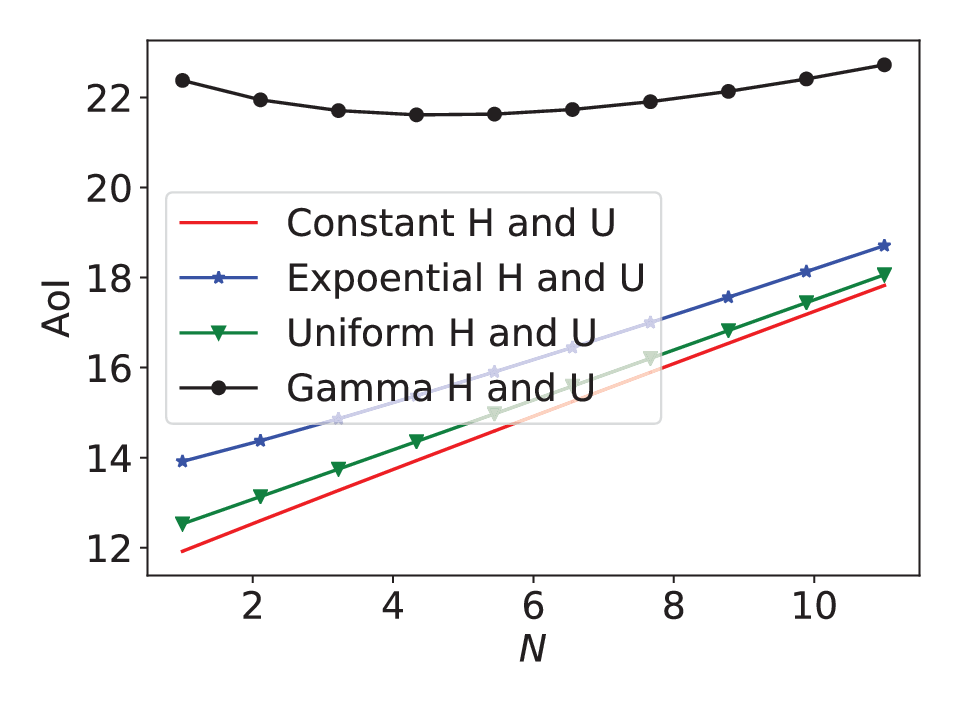}}
\subfloat[Single-sleep scheme]{\includegraphics[scale=0.35]{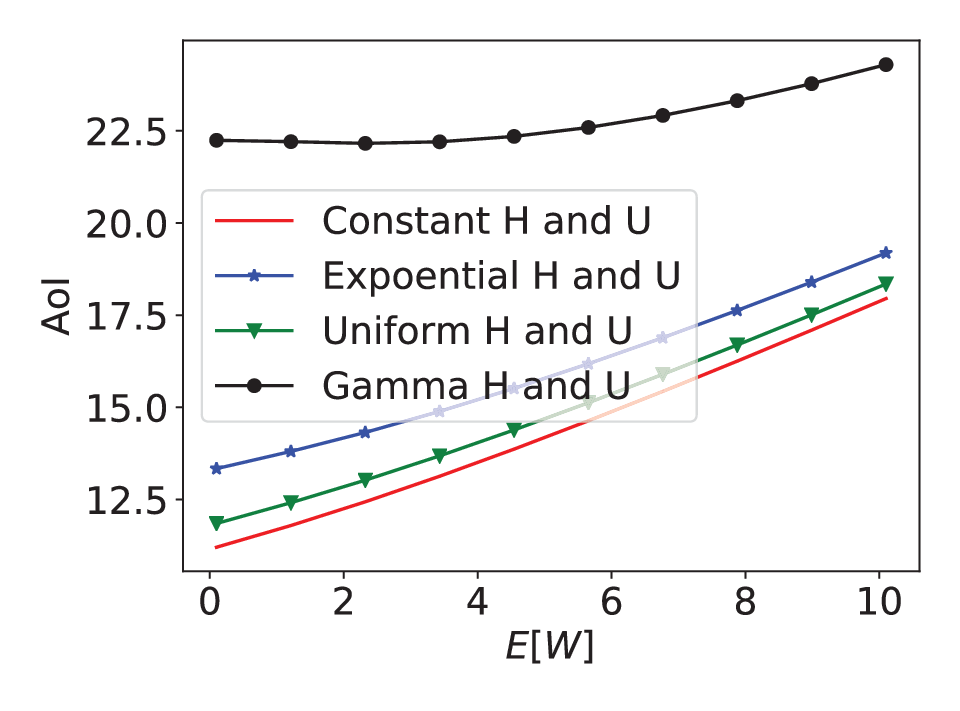}}
\subfloat[Multiple-sleep scheme]{\includegraphics[scale=0.35]{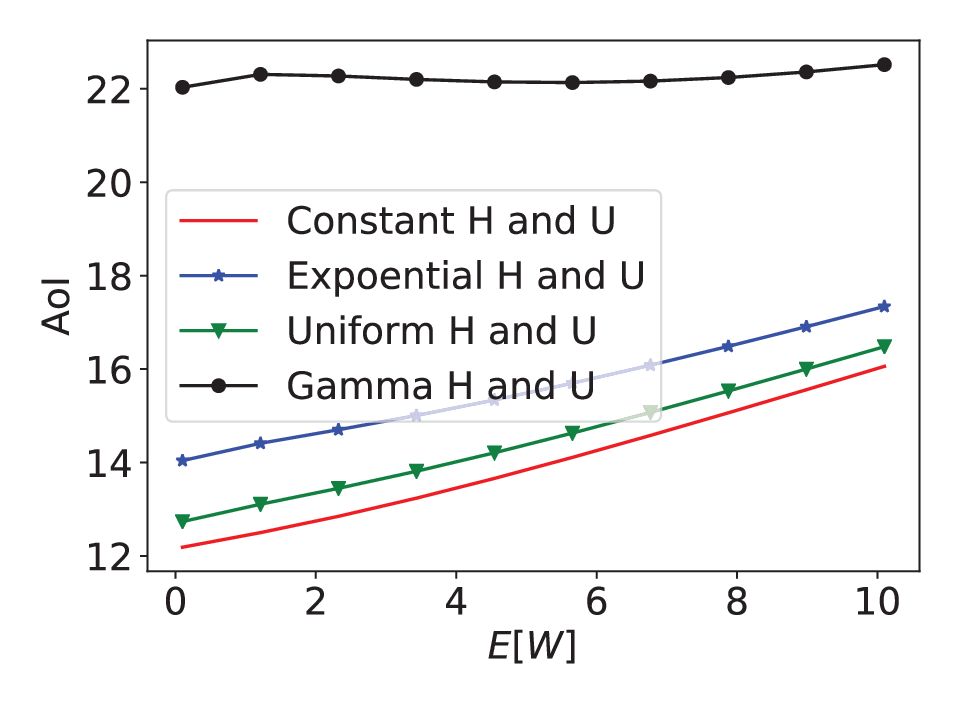}}

\caption{AoI under CS with different service and setup time distributions.
$\lambda=0.8$, $\boldsymbol{E}[H]=5$, $\boldsymbol{E}[U]=5$. \label{fig:AoI-Under-Different}}
\end{figure*}

In Fig. \ref{fig:Age-energy-Tradeoff}, we plot the energy consumption
rate, PAoI, and AoI as functions of the variable $b$ and $N$ for
the single-source scenario. We can develop several insights from Fig.
\ref{fig:Age-energy-Tradeoff}. First, we observe from Fig. \ref{fig:Age-energy-Tradeoff}(a)
that the energy consumption rate is always a decreasing function of
$N$, which means sleeping for a long time would reduce the average
energy consumption. Second, from Fig. \ref{fig:Age-energy-Tradeoff}(a)
we observe that the energy consumption could be a decreasing or increasing
function of the threshold $b$, as we proved in Corollary \ref{cor:When-,-then}.
Third, we observe from Fig. \ref{fig:Age-energy-Tradeoff}(b) and
Fig. \ref{fig:Age-energy-Tradeoff}(c) that the shapes of the PAoI
and AoI functions are distinct. The PAoI declines if $N$ decreases
or $b$ increases, meaning that sleeping for a shorter period or less
frequently can reduce PAoI. In contrast, as shown in Fig. \ref{fig:Age-energy-Tradeoff}(c),
enlarging the sleeping period or probability does not always increase
the AoI.

In Fig. \ref{fig:AoI-Under-Different}(a), we plot the AoI under N-policy
with $b=0$. It can be observed that AoI is an increasing function
of $N$ when $H$ and $U$ are both constants ($CV=0$), exponential
($CV=1$), and uniform ($CV=\frac{\sqrt{3}}{3}$). When $H\sim Gamma(\frac{1}{5},25)$
and $U\sim Gamma(\frac{1}{5},25)$, we have $CV$ for $H$ and $U$
are both $\sqrt{5}$. As shown in Fig. \ref{fig:AoI-Under-Different}(a),
the optimal $N^{*}=4$ in this case. 

We now numerically show that this non-monotonicity is not due to using
N-policy as the wakeup scheme. In Fig. \ref{fig:AoI-Under-Different}(b)
and \ref{fig:AoI-Under-Different}(c), we plot the AoI under the single-sleep
and multiple-sleep schemes, and we let the sleeping period $W$ be
exponential for both the single-sleep and multiple-sleep schemes.
The sleeping span within each regenerative cycle under the single-sleep
scheme is thus $\boldsymbol{E}[W]$, and that under the multiple-sleep
scheme is $\boldsymbol{E}[W]+\frac{1}{\lambda}$ (see Appendix \ref{sec:Performance-Metrics-for}).
From Fig. \ref{fig:AoI-Under-Different}(b) and \ref{fig:AoI-Under-Different}(c),
we see that the AoI is not monotone on the value of $\boldsymbol{E}[W]$
when both $H$ and $U$ are gamma-distributed. This observation means
that increasing sleeping length does not always increase AoI, and
this phenomenon is not unique to using N-policy as the wakeup scheme.
Moreover, we show in Appendix \ref{sec:Discussion-on-the} of the
supplementary material that this phenomenon also exists in the M/G/1/LCFS
system. 

\subsection{Comparison between Infinite and Bufferless Systems\label{subsec:Comparison-between-LCFS}}

In the bufferless system, the server does not need to process all
the data packets due to some packets being dropped when the server
is processing, sleeping, or setting up. We now show that the bufferless
system thus achieves a better PAoI-energy tradeoff than the infinite
buffer system when the sampling rate is large.

\begin{figure}
\begin{center}\includegraphics[scale=0.35]{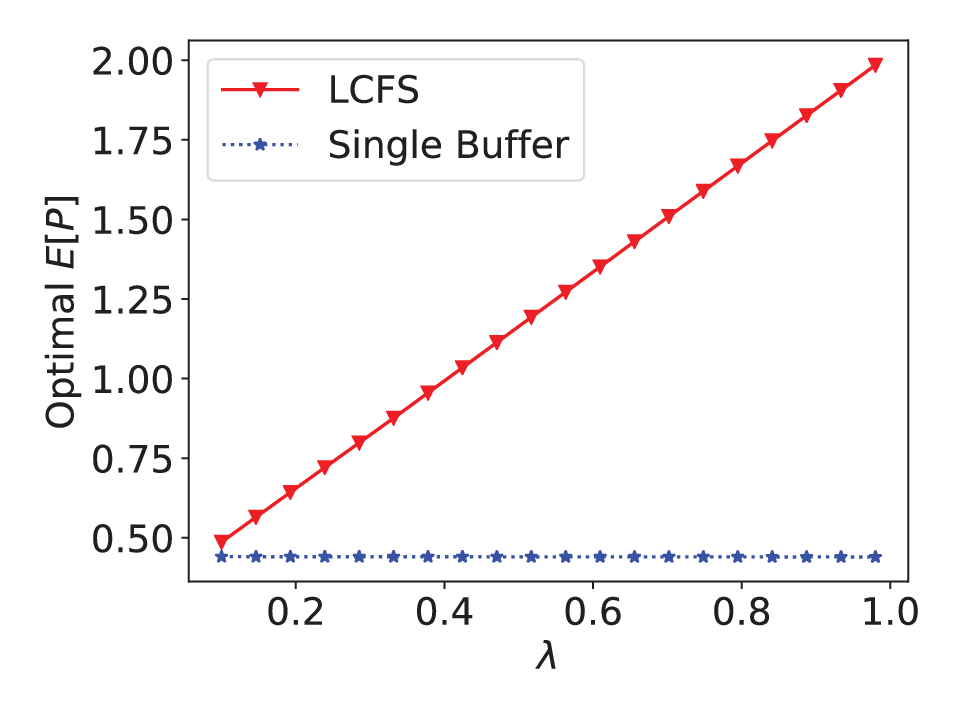}\end{center}
\caption{\label{fig:Optimal-Energy-Consumption}Optimal energy consumption
for LCFS and bufferless systems. $H\sim exp(1),$ $U=0.5$, $P_{B}=2.1,$ $P_{ID}=1.1,$ $P_{SL}=0.3,$ $P_{ST}=0.5,$
$\tau=15$}
\end{figure}

The PAoI for a single source system with the infinite buffer size,
LCFS service discipline, HT as the idling scheme, and N-policy as
the wakeup scheme was investigated in \cite{xu2021Information}. To
make a fair comparison, we consider a single-source LCFS system with
CS as the idling scheme and N-policy as the sleeping scheme. Based
on the analysis in \cite{xu2021Information} and our discussion in
Section \ref{sec:Queueing-Analysis}, we can derive its PAoI as
\begin{align}
& \boldsymbol{E}[A_{LCFS}^{CS}] =  \boldsymbol{E}[H]+\bigg[\frac{\theta(1-\lambda\boldsymbol{E}[H])(1-U^{*}(\lambda))}{1+\theta(N-1+\lambda\boldsymbol{E}[U])}\nonumber\\
&+2-H^{*}(\lambda)+\lambda H^{*(1)}(\lambda)\bigg]\nonumber\\
  & \bigg/\lambda\bigg[-H^{*}(\lambda)+1+\frac{1-\lambda\boldsymbol{E}[H]}{1+\theta(N-1+\lambda\boldsymbol{E}[U])}\bigg],\label{eq:13-1}
\end{align}
and the energy consumption rate as 
%\boldsymbol{E}[P_{LCFS}^{CS}] = & \lambda\boldsymbol{E}[H]P_{B}+\bigg(1-\lambda\boldsymbol{E}[H]\bigg)\bigg(\frac{(1-\theta)P_{ID}}{\lambda}\nonumber \\
%  & +\theta\frac{N}{\lambda}P_{SL}+\theta\boldsymbol{E}[U]P_{ST}\bigg)\bigg/\bigg[\frac{1-\theta}{\lambda}+\theta(\frac{N}{\lambda}+\boldsymbol{E}[U])\bigg].\label{eq:13-2}
\begin{align}
& \boldsymbol{E}[P_{LCFS}^{CS}] =  \lambda\boldsymbol{E}[H]P_{B}+\bigg(1-\lambda\boldsymbol{E}[H]\bigg)\bigg(\frac{(1-\theta)P_{ID}}{\lambda}\nonumber \\
 & +\theta\frac{N}{\lambda}P_{SL}+\theta\boldsymbol{E}[U]P_{ST}\bigg)\bigg/\bigg[\frac{1-\theta}{\lambda}+\theta(\frac{N}{\lambda}+\boldsymbol{E}[U])\bigg].\label{eq:13-2}
\end{align}
The derivation of the sleeping probability $\theta$ is left to Appendix
\ref{sec:Discussion-on-the} of the supplementary material. We compare
the minimal energy consumption rate under LCFS and bufferless system
when both of their PAoI is constrained by the same constant in Fig.
\ref{fig:Optimal-Energy-Consumption}. We observe from Fig. \ref{fig:Optimal-Energy-Consumption}
that the energy consumption under the bufferless system is much lower
than LCFS when the packet generation rate is large. The reason is
that under LCFS, the server has to process all the arrived data packets.
From Equations (\ref{eq:13-1}) and (\ref{eq:13-2}) we can find that
when the sampling rate is large, altering the value of $\theta$ and
$N$ under LCFS does not change PAoI and energy consumption significantly,
as the server will spend most of the time processing. Therefore, using
the bufferless system can achieve a better PAoI-energy tradeoff when
the sampling rate is large. 

\section{Conclusion and Future Research\label{sec:Conclusion-and-Future}}

In this paper, we investigate the information freshness and energy
consumption in a single server queueing system where the server can
sleep to reduce its energy consumption. We propose a modeling approach
that relies on a renewal-type analysis to derive the closed-form expressions
for information metrics (i.e., PAoI and AoI) and energy consumption
rate. Furthermore, we propose an idling scheme called Conditional
Sleep (CS) scheme to achieve the same PAoI and energy consumption
rate as two widely used strategies (namely Hysteresis Time (HT) scheme
and Bernoulli Sleep (BS) scheme) while achieving a smaller AoI than
these two policies.

We analytically find that extending the server's sleeping period length
can reduce energy consumption and enlarge the PAoI, but not always
increase the AoI. We derive the conditions under which increasing
sleeping period length does not increase AoI. We show that this counter-intuitive
phenomenon occurs when the packet processing time and setup time distributions
have a large coefficient of variation. Our analysis further shows
that optimizing the energy consumption under a PAoI constraint can
result in an optimal solution that is difficult to implement in practice,
but this issue does not exist when using AoI in the information freshness
constraint. We also find that using the bufferless system can achieve
a better PAoI-energy tradeoff than having an infinite buffer size
when the packet arrival rate is large.

In this research, we mainly consider the scheduling schemes from the
server's perspective. We will consider the cooperation between data
sources and the server in our future research. Moreover, we hope to
extend our discussion to the scenario with multiple servers coordinating
sleep-wake strategies. 

\begin{appendices}

\section{\label{sec:Derivations-of-the}Proof of Theorem \ref{thm:(1)-Under-HT,}}

Since the derivations for HT, BS, and CS are similar and only differ
in the LST of the regeneration cycles, we first introduce the derivation
framework for HT in detail in Appendix \ref{subsec:Hysteresis-Time-Strategy},
and use the same framework to derive for BS and CS in Appendices \ref{subsec:Bernoulli-Sleep-Strategy}
and \ref{subsec:Conditional-Sleep-Scheme}.

\subsection{Hysteresis Time Scheme\label{subsec:Hysteresis-Time-Strategy}}

To derive the closed-form expressions for system performance metrics
for HT, we first derive the LST of regenerative cycles. We then derive
the closed-form expression of energy consumption and information freshness
metrics based on the LST of regenerative cycles.

\subsubsection{Regenerative Cycles}

As we defined in Section \ref{sec:Queueing-Analysis}, a Class $i$
regenerative cycle $V_{i}$ starts from processing a packet from data
source $i$, and ends when the server starts processing a packet.
After processing a packet, the server under HT will remain to idle
until either of the following two cases occurs: 1) an arrival occurs
before the hysteresis time $D_{i}$ is over, or 2) it has idled for
time $D_{i}$. In the first case, $V_{i}$ is over when a new arrival
occurs. Let $L$ be the inter-arrival time of data packets. Following
the superposition of Poisson arrival processes (see \cite{kulkarni2016modeling})
and the memoryless property of exponential inter-arrival times of
packets, we obtain the LST of the idling period as
\begin{eqnarray*}
 &  & \boldsymbol{E}[e^{-sL}|D_{i}\geq L]\boldsymbol{P}(D_{i}\geq L)=\frac{\lambda}{\lambda+s}(1-D_{i}^{*}(s+\lambda)).
\end{eqnarray*}
The LST of $V_{i}$ in this case is then given by $H_{i}^{*}(s)\frac{\lambda}{\lambda+s}(1-D_{i}^{*}(s+\lambda))$.
In the second case, no arrival occurs during $D_{i}$. The period
$V_{i}$ will contain an idling period, a sleeping period, and a setup
period. The idling period's LST, in this case, is 
\begin{eqnarray*}
 &  & \boldsymbol{E}[e^{-sD_{i}}|D_{i}\leq L]\boldsymbol{P}(D_{i}\leq L)=D_{i}^{*}(\lambda+s).
\end{eqnarray*}
After the idling period, the server will further experience a sleeping
period until $N$ packets have arrived, and a setup period $U$. So
the LST of $V_{i}$ in the second case is $H_{i}^{*}(s)D_{i}^{*}(\lambda+s)(\frac{\lambda}{\lambda+s})^{N}U^{*}(s)$.
By combining the LST of $V_{i}$ in the above two cases, we have
\begin{eqnarray}
V_{i}^{*}(s) & =&  H_{i}^{*}(s)\bigg[\frac{\lambda}{s+\lambda}(1-D_{i}^{*}(s+\lambda))\nonumber\\
& & +D_{i}^{*}(s+\lambda)(\frac{\lambda}{\lambda+s})^{N}U^{*}(s)\bigg].\label{eq:1-1}
\end{eqnarray}
Moreover, we can derive the probability that the server sleeps within
a Class $i$ regenerative cycle as $\theta_{i}^{HT}=\boldsymbol{P}(D_{i}\leq L)=D_{i}^{*}(\lambda).$
We will rely on the closed-form expressions of $V_{i}^{*}(s)$ and
$\theta_{i}^{HT}$ in our derivations later.

\subsubsection{Energy Consumption Rate}

We now use the results of regenerative cycles to derive the energy
consumption rate. Notice that each regenerative cycle starts by processing
a packet, and this data packet should be either 1) the data packet
that arrived during the idling period of the previous regenerative
cycle, or 2) the last packet that arrived during the sleeping or setup
period in the previous regenerative cycle. By the Bernoulli splitting
of Poisson processes \cite{kulkarni2016modeling}, this packet has
probability $\frac{\lambda_{i}}{\lambda}$ to belong to data source
$i$. As a result, the probability of having a Class $i$ regenerative
cycle is $\frac{\lambda_{i}}{\lambda}$. The expected length for Class
$i$ regenerative cycle is $\boldsymbol{E}[V_{i}]=-V_{i}^{*(1)}(0)$
, then the expected length of a regenerative cycle is given as 
\begin{align}
\sum_{i=1}^{k}\frac{\lambda_{i}}{\lambda}\boldsymbol{E}[V_{i}]= & \sum_{i=1}^{k}\frac{\lambda_{i}}{\lambda}\bigg[\boldsymbol{E}[H_{i}]+\theta_{i}^{HT}(\frac{N-1}{\lambda}+\boldsymbol{E}[U])+\frac{1}{\lambda}\bigg].
\end{align}
Letting $\lambda_{e}$ be the arrival rate of the regenerative cycles,
and from the fact that $\lambda_{e}\sum_{i=1}^{k}\frac{\lambda_{i}}{\lambda}\boldsymbol{E}[V_{i}]=1$,
we have
\begin{equation}
\lambda_{e}=1\bigg/\bigg\{\sum_{i=1}^{k}\frac{\lambda_{i}}{\lambda}\big[\boldsymbol{E}[H_{i}]+\theta_{i}^{HT}(\frac{N-1}{\lambda}+\boldsymbol{E}[U])+\frac{1}{\lambda}\big]\bigg\}.
\end{equation}
Then the expected energy consumption rate is given by 
\begin{align}
\boldsymbol{E}[P^{HT}] & =\lambda_{e}\sum_{i=1}^{k}\frac{\lambda_{i}}{\lambda}\bigg[P_{B}\boldsymbol{E}[H_{i}]+\frac{1-\theta_{i}^{HT}}{\lambda}P_{ID}\nonumber\\
&+\theta_{i}^{HT}(\frac{N}{\lambda}P_{SL}+\boldsymbol{E}[U]P_{ST})\bigg].\label{eq:1}
\end{align}

\subsubsection{Information Freshness}

We now introduce the way to derive AoI and PAoI. In each regenerative
cycle, the server only processes one packet. We call these packets
that finish processing the ``informative packets''. Note that only
informative packets result in age drops (see \cite{costa2016age}).
We let $G_{i}$ denote the waiting time of an informative packet from
data source $i$, and $I_{ii}$ be the time span from processing a
packet from source $i$, to the next time the server is about to process
a packet from source $i$. Then we can derive the LST of age peak
$A_{i}^{*}(s)$ for user $i$ as 
\begin{eqnarray}
A_{i}^{*}(s) & = & G_{i}^{*}(s)I_{ii}^{*}(s)H_{i}^{*}(s).\label{eq:2}
\end{eqnarray}
\begin{figure}
\center{\includegraphics[scale=0.3]{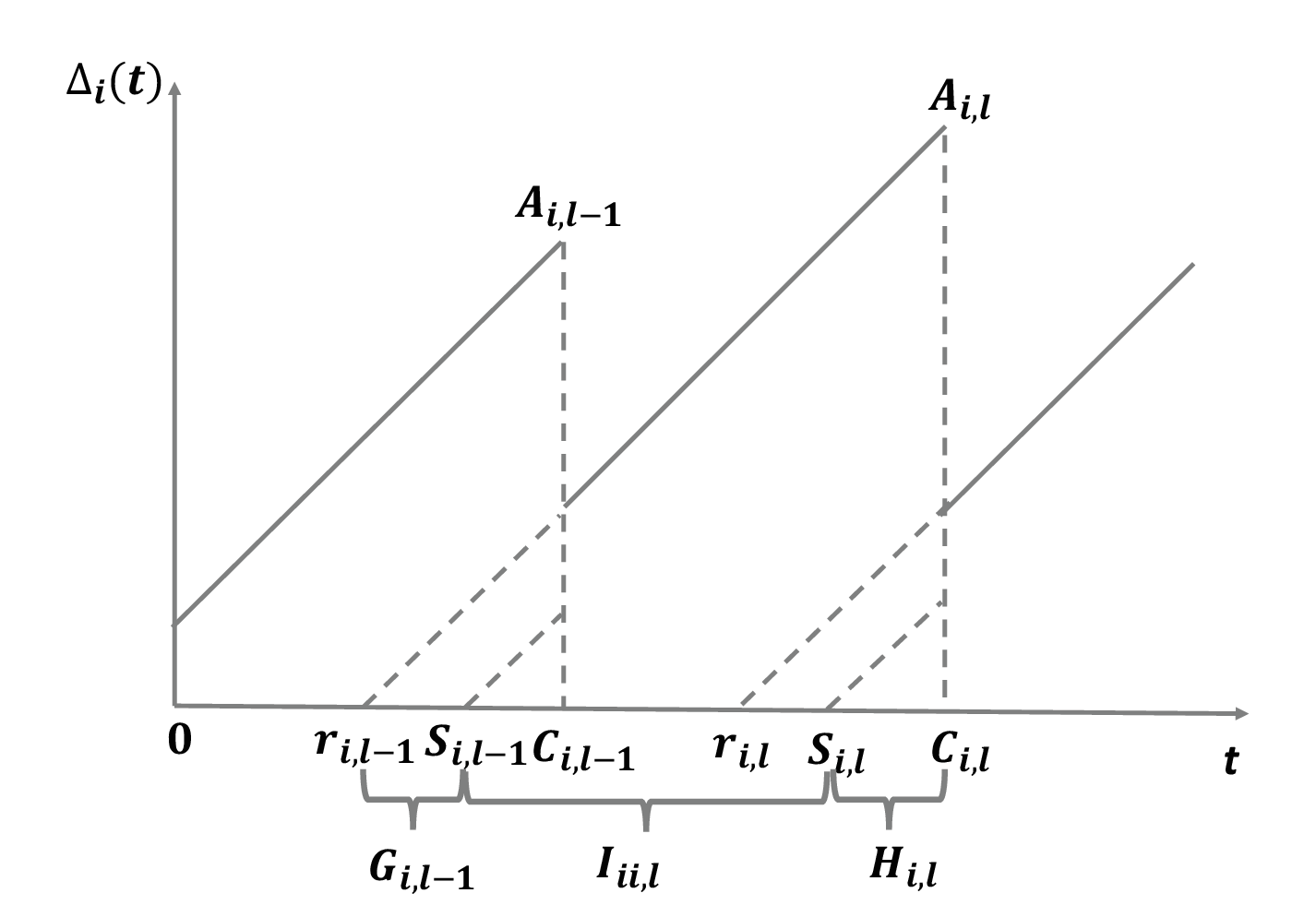}}
\caption{Age of Information process\label{fig:Information-Freshness}. The
variables $r_{i,l}$, $S_{i,l}$, and $C_{i,l}$ are generation time,
processing starting time, and completion time of the $l^{th}$ informative
packet for data source $i$. $G_{i,l}$, $I_{ii,l}$, $H_{i,l}$,
and $A_{i,l}$ are the waiting time, inter-processing time, processing
time, and age peak for the $l^{th}$ informative packet. }
\end{figure}
Equation (\ref{eq:2}) holds because the peak age occurs only before
an informative packet completes the service. Before the packet completes
service, the age at that time instance is equal to the period from
the generation time of the previous informative packet from source
$i$ to the current time. This time span is comprised of the waiting
time $G_{i}$ of the previous packet, period $I_{ii}$, and the processing
time $H_{i}$ of the current packet. The three components $G_{i}$,
$I_{ii}$, and $H_{i}$ are mutually independent. The reason is that
the waiting time of the previous packet $G_{i}$ is independent of
the time span $I_{ii}$ and the processing time $H_{i}$ of the current
packet. The processing time $H_{i}$ of the current packet is independent
of the period $I_{ii}$. Fig. \ref{fig:Information-Freshness} provides
a demonstrative graph for the age process. By a similar argument to
Eq. (1) and (2) in \cite{xu2020vacations}, the expected PAoI for
user $i$ can be given as
\begin{eqnarray}
\boldsymbol{E}[A_{i}] & = & \boldsymbol{E}[G_{i}]+\boldsymbol{E}[I_{ii}]+\boldsymbol{E}[H_{i}],\label{eq:3}
\end{eqnarray}
and the AoI for data source $i$ is 
\begin{eqnarray}
\boldsymbol{E}[\Delta_{i}] & = & \frac{\boldsymbol{E}[I_{ii}^{2}]}{2\boldsymbol{E}[I_{ii}]}+\boldsymbol{E}[G_{i}]+\boldsymbol{E}[H_{i}].\label{eq:4}
\end{eqnarray}
We now derive $\boldsymbol{E}[G_{i}]$. An informative packet does not need to wait if it arrives during
an idling period. It will wait if it arrives when the server
is setting up. So the informative packet is either the $N^{th}$ packet
that arrives during the sleeping period, or the last packet that arrives
during $U$. The probability that the server experiences a setup period
within a regenerative cycle is $\sum_{i=1}^{k}\theta_{i}^{HT}\frac{\lambda_{i}}{\lambda}$.
Since the expected waiting time of an informative packet does not
depend on the data source that it comes from, using Lemma 1 of \cite{xu2019towards},
we have 
\begin{eqnarray}
\boldsymbol{E}[G_{i}]  =  \sum_{i=1}^{k}\theta_{i}^{HT}\frac{\lambda_{i}}{\lambda}\frac{1-U^{*}(\lambda)}{\lambda}.\label{eq:4-2}
\end{eqnarray}
To obtain $\boldsymbol{E}[A_{i}]$ and $\boldsymbol{E}[\Delta_{i}]$,
we further need to obtain the first and second moment of $I_{ii}$.
We define $I_{ji}$ as the time span from processing a source $j$
packet, to the next time a source $i$ packet starts service. We then
have 
\begin{eqnarray}
 &  & I_{ji}^{*}(s)=V_{j}^{*}(s)\bigg[\sum_{l\neq i}\frac{\lambda_{l}}{\lambda}I_{li}^{*}(s)+\frac{\lambda_{i}}{\lambda}\bigg],\nonumber\\
 &  & I_{ii}^{*}(s)=\frac{V_{i}^{*}(s)\frac{\lambda_{i}}{\lambda}}{1-\sum_{l\neq i}\frac{\lambda_{l}}{\lambda}V_{l}^{*}(s)}.\label{eq:5}
\end{eqnarray}
By Equations (\ref{eq:1-1}), (\ref{eq:3}), (\ref{eq:4}), (\ref{eq:4-2}),
and (\ref{eq:5}), we can then obtain the AoI and PAoI under HT.

\subsection{Bernoulli Sleep Scheme\label{subsec:Bernoulli-Sleep-Strategy}}

Under BS, the server sleeps with probability $\theta_{i}^{BS}$ after
processing a packet from source $i$. Following \cite{wu2013single,maraghi2009batch},
BS is easy to understand and implement in practice. Similar to the
discussion of HT, the expected length of each regenerative cycle under
BS is 
\begin{eqnarray}
\sum_{i=1}^{k}\frac{\lambda_{i}}{\lambda}\bigg[\boldsymbol{E}[H_{i}]+\theta_{i}^{BS}(\frac{N-1}{\lambda}+\boldsymbol{E}[U])+\frac{1}{\lambda}\bigg].
\end{eqnarray}
From the same argument for HT, the expected energy consumption rate
under BS then has the same expression as that under HT. We then derive
the AoI and PAoI under BS. Similar to HT, the waiting time of informative
packets under BS only appears when the server is sleeping or setting
up, thus $\boldsymbol{E}[G_{i}]=\sum_{i=1}^{k}\theta_{i}^{BS}\frac{\lambda_{i}}{\lambda}\frac{1-U^{*}(\lambda)}{\lambda}.$
Under BS, the LST of a Class $i$ regenerative cycle is given as
\begin{align}
V_{i}^{*}(s)=H_{i}^{*}(s)\big[\frac{\lambda}{s+\lambda}(1-\theta_{i}^{BS})+\theta_{i}^{BS}(\frac{\lambda}{\lambda+s})^{N}U^{*}(s)\big].\label{eq:7-1}
\end{align}
By Equations (\ref{eq:3})-(\ref{eq:5}) and (\ref{eq:7-1}), we can
then obtain the AoI and PAoI under BS.

\subsection{Conditional Sleep Scheme\label{subsec:Conditional-Sleep-Scheme}}

The server under CS only sleeps when the service time $H_{i}$ is
smaller than the threshold variable $B_{i}$. The idea of CS is to
remain idling after processing a packet for a long time, so that to
reduce the peak age of the next regenerative cycle. When \textbf{$B_{i}$
}is exponentially distributed with rate $b_{i}$, the sleeping probability
after serving a source $i$ packet is thus given as $\theta_{i}^{CS}=\boldsymbol{P}(H_{i}<B_{i})=H_{i}^{*}(b_{i}).$
From the same argument for HT, the energy consumption rate under CS
then have the same expression as that under HT. From the similar argument
for BS, under CS we have $\boldsymbol{E}[G_{i}]=\sum_{i=1}^{k}\theta_{i}^{CS}\frac{\lambda_{i}}{\lambda}\frac{1-U^{*}(\lambda)}{\lambda}.$
The LST of Class $i$ regenerative cycle under CS is then given by
\begin{align}
V_{i}^{*}(s)  = & \boldsymbol{E}[e^{-sH_{i}}|H_{i}>B_{i}]\frac{\lambda}{s+\lambda}\boldsymbol{P}(H_{i}>B_{i})\nonumber \\
  & +\boldsymbol{E}[e^{-sH_{i}}|H_{i}\leq B_{i}](\frac{\lambda}{\lambda+s})^{N}U^{*}(s)\boldsymbol{P}(H_{i}\leq B_{i})\nonumber \\
  = & \big[H_{i}^{*}(s)-H_{i}^{*}(s+b)\big]\frac{\lambda}{s+\lambda}\nonumber\\
  & +H_{i}^{*}(s+b)(\frac{\lambda}{\lambda+s})^{N}U^{*}(s).\label{eq:8-1}
\end{align}
By Equations (\ref{eq:3})-(\ref{eq:5}), and (\ref{eq:8-1}), we
can compute the AoI and PAoI under CS and derive the expressions
in Theorem \ref{thm:(1)-Under-HT,}.

\section{Proof of Theorem \ref{thm:If--is-1}\label{sec:Proof-of-Theorem}}
\begin{IEEEproof}
When $\theta_{1}^{HT}=\theta_{1}^{BS}=\theta_{1}^{CS}$, then HT,
BS, and CS have the same energy consumption and PAoI. We first show
that $\boldsymbol{E}[\Delta_{1}^{BS}]\leq\boldsymbol{E}[\Delta_{1}^{HT}]$.
When $\theta_{1}^{BS}=\theta_{1}^{HT}$, then have the same $\boldsymbol{E}[I_{11}]$,
$\boldsymbol{E}[H_{1}]$, $H_{1}^{*(2)}(0)$, and $\boldsymbol{E}[G]$
for BS and HT. Therefore 
\begin{eqnarray*}
 &  & \boldsymbol{E}[\Delta_{1}^{BS}]-\boldsymbol{E}[\Delta_{1}^{HT}]\\
 & = & -\frac{\frac{2}{\lambda}D_{1}^{*(1)}(\lambda_{1})-2D_{1}^{*(1)}(\lambda_{1})\frac{N}{\lambda}-2D_{1}^{*(1)}(\lambda_{1})\boldsymbol{E}[U]}{2\boldsymbol{E}[I_{11}]}\\
 & = & -D_{1}^{*(1)}(\lambda_{1})\frac{\frac{1}{\lambda_{1}}(1-N)-\boldsymbol{E}[U]}{\boldsymbol{E}[I_{11}]}\leq0.
\end{eqnarray*}
We now show $\boldsymbol{E}[\Delta_{1}^{CS}]\leq\boldsymbol{E}[\Delta_{1}^{BS}]$.
Since BS and CS have the same $\boldsymbol{E}[I_{11}]$, $\boldsymbol{E}[H_{1}]$,
$H_{1}^{*(2)}(0)$, and $\boldsymbol{E}[G_{1}]$, we then have 
\begin{eqnarray*}
& & \boldsymbol{E}[\Delta_{1}^{CS}]-\boldsymbol{E}[\Delta_{1}^{BS}] \nonumber \\
& = & -\frac{(\boldsymbol{E}[H_{1}]\boldsymbol{E}[e^{-b_{1}H_{1}}]-\boldsymbol{E}[H_{1}e^{-b_{1}H_{1}}])(\frac{N-1}{\lambda_{1}}+\boldsymbol{E}[U])}{\boldsymbol{E}[I_{11}]}\nonumber\\
&\leq& 0,
\end{eqnarray*}
with the last inequality following from the fact that $\boldsymbol{E}[H_{1}]\boldsymbol{E}[e^{-b_{1}H_{1}}]-\boldsymbol{E}[H_{1}e^{-b_{1}H_{1}}]\geq0$
(see \cite{lehmann1966some,xu2020vacations}).
\end{IEEEproof}

\small{\bibliographystyle{IEEEtran}
\bibliography{Jin_AoI_2}
}\normalsize{}

\newpage
\pagebreak
$ $
\newpage
\begin{strip}
\begin{spacing}{1.0}
\begin{center} \LARGE{\textbf{Supplementary Material for the Paper ``How should the server sleep? -- Age-energy tradeoff in sleep-wake server systems''}} \end{center}
\end{spacing}
\end{strip}

\section{\label{sec:Performance-Metrics-for}Proof of Proposition \ref{prop:When-fixing-CS}}

We here only derive the performance metrics for single-sleep and multiple-sleep
strategies under CS. The derivation processes for single-sleep and
multiple-sleep schemes under BS and HT are similar, and we omit them
here. 

\subsection{Single-sleep}

Under the single-sleep scheme, the server sleeps for a random period
$W$, and then sets up. If no arrival occurs during the sleeping period
or the setup period, the server will stay idling after setting up
until a new arrival occurs. If there is a packet waiting in the system
when the setup period is over, then the server will process the packet
immediately. Still, we only keep the last packet if multiple packets
arrive during the setup period.

The energy consumption rate can be derived by considering the events
within each regenerative cycle. Since the probability that no arrival
occurs during the sleep period nor the setup period is given by $W^{*}(\lambda)U^{*}(\lambda)$,
we then have 
\begin{align}
&\boldsymbol{E}[P^{CS}] =\nonumber\\
 & \bigg\{\sum_{i=1}^{k}\frac{\lambda_{i}}{\lambda}\bigg[P_{B}\boldsymbol{E}[H_{i}]+\frac{1-\theta_{i}^{CS}}{\lambda}P_{ID}\nonumber\\
&+\theta_{i}^{CS}(\boldsymbol{E}[W]P_{SL}+\boldsymbol{E}[U]P_{ST}+\frac{W^{*}(\lambda)U^{*}(\lambda)}{\lambda}P_{ID})\bigg]\bigg\}\nonumber \\
 & \bigg/\bigg\{\sum_{i=1}^{k}\frac{\lambda_{i}}{\lambda}\bigg[\boldsymbol{E}[H_{i}]+\theta_{i}^{CS}(\boldsymbol{E}[W]+\boldsymbol{E}[U]\nonumber\\
 &+\frac{W^{*}(\lambda)U^{*}(\lambda)}{\lambda})+\frac{1-\theta_{i}^{CS}}{\lambda}\bigg]\bigg\}.\label{eq:10-1}
\end{align}
Note that the $P_{SL}$ under single-sleep could be smaller than the
$P_{SL}$ under N-policy, since under single-sleep, the server does
not incur a cost for detecting and counting the number of arrivals
during the sleeping period (see \cite{guo2016delay}). 

To derive the AoI and PAoI, we only need to characterize the LST of
regenerative cycles (i.e., $V_{i}^{*}(s)$) and the expectation of
informative packets' waiting time (i.e., $\boldsymbol{E}[G_{i}]$),
then apply Equations (\ref{eq:3}), (\ref{eq:4}), and (\ref{eq:5}).
Conditioning on the cases of whether the inter-arrival time of packets
$L$ satisfies $L>W+U$, the expression of $V_{i}^{*}(s)$ under single-sleep
scheme 
\begin{align}
&V_{i}^{*}(s)  =\bigg[H_{i}^{*}(s)-H_{i}^{*}(s+b)\bigg]\frac{\lambda}{s+\lambda}\nonumber \\
 & +H_{i}^{*}(s+b)\bigg[W^{*}(s)U^{*}(s)-\frac{s}{s+\lambda}W^{*}(s+\lambda)U^{*}(s+\lambda)\bigg].\label{eq:10-3}
\end{align}

We now introduce the way to calculate $\boldsymbol{E}[G_{i}]$. Since
an informative packet only has to wait if it arrives when the server
is sleeping or setting up and the probability that the server experiences
a sleeping/setup period is $\sum_{i=1}^{k}\theta_{i}^{CS}\frac{\lambda_{i}}{\lambda}$,
using the derivation for single-sleep scheme in \cite{xu2021Information},
we have 
\begin{eqnarray}
\boldsymbol{E}[G_{i}] & = & \sum_{j=1}^{k}\theta_{j}^{CS}\frac{\lambda_{j}}{\lambda}(\frac{1-W^{*}(\lambda)U^{*}(\lambda)}{\lambda}\nonumber\\
& & +W^{*(1)}(\lambda)U^{*}(\lambda)+W^{*}(\lambda)U^{*(1)}(\lambda)).\label{eq:10-4}
\end{eqnarray}
Then, we can compute the PAoI and AoI by combing Equations (\ref{eq:3})-(\ref{eq:5}),
(\ref{eq:10-3}), and (\ref{eq:10-4}). We do not present the closed-form
expressions of PAoI and AoI here as they are involved.

\subsection{Multiple-sleep}

Under the multiple-sleep scheme, if the server returns from a sleeping
period $W$ and finds the system non-empty, then the server wakes
up; otherwise, another sleeping period $W$ is taken. We assume that
only the freshest packet is kept in the system when the server is
sleeping or setup. The energy consumption rate $P_{SL}$ under the
multiple-sleep scheme is also lower than that under N-policy, as the
multiple-sleep scheme does not need to count the arrival packets.
However, there may exist a detection cost $P_{DT}$ whenever the server
returns from a sleeping period to detect whether packets are waiting
(see \cite{guo2016delay}). 

From \cite{xu2020vacations,xu2021Information}, we know that the averaged
sleeping length under multiple-sleep scheme is $\frac{\boldsymbol{E}[W]}{1-W^{*}(\lambda)}$,
and the number of sleeping periods that the server has before setting
up is $\frac{1}{1-W^{*}(\lambda)}$. We then have 
\begin{align}
& \boldsymbol{E}[P^{CS}]=  \bigg\{\sum_{i=1}^{k}\frac{\lambda_{i}}{\lambda}\big[P_{B}\boldsymbol{E}[H_{i}]+\frac{1-\theta_{i}^{CS}}{\lambda}P_{ID}\nonumber\\
&+\theta_{i}^{CS}\big(\frac{\boldsymbol{E}[W]P_{SL}}{1-W^{*}(\lambda)}+\frac{P_{DT}}{1-W^{*}(\lambda)}+\boldsymbol{E}[U]P_{ST}\big)\big]\bigg\}\nonumber \\
 & \big/\bigg\{\sum_{i=1}^{k}\frac{\lambda_{i}}{\lambda}\big[\boldsymbol{E}[H_{i}]+\theta_{i}^{CS}(\frac{\boldsymbol{E}[W]}{1-W^{*}(\lambda)}+\boldsymbol{E}[U])+\frac{1-\theta_{i}^{CS}}{\lambda}\big]\bigg\}.\label{eq:10-2}
\end{align}
Using the LST of vacation period derived in \cite{xu2020vacations},
we have 
\begin{align}
V_{i}^{*}(s) & =\bigg[H_{i}^{*}(s)-H_{i}^{*}(s+b)\bigg]\frac{\lambda}{s+\lambda}\nonumber\\
&+H_{i}^{*}(s+b)\frac{W^{*}(s)-W^{*}(s+\lambda)}{1-W^{*}(s+\lambda)}U^{*}(s),\label{eq:10-5}
\end{align}
and the expected waiting time for informative packets is given by
\begin{equation}
\boldsymbol{E}[G_{i}]=\sum_{i=1}^{k}\theta_{i}^{CS}\frac{\lambda_{i}}{\lambda}(\frac{1}{\lambda}+U^{*}(\lambda)\frac{W^{*(1)}(\lambda)}{1-W^{*}(\lambda)}).\label{eq:10-6}
\end{equation}
We can then obtain the PAoI and AoI using Equations (\ref{eq:3})-(\ref{eq:5}),
(\ref{eq:10-5}), and (\ref{eq:10-6}). Table \ref{tab:Simulation-Results-v.s.-1}
provides a comparison of the simulation results and analytical results.
We can see that the simulation results match the analytical results,
which verify our derivations.

\begin{table*}
\footnotesize{\begin{center}%
\begin{tabular}{|>{\raggedright}p{1.2cm}|c|c|c|c|c|}
\hline 
\multicolumn{2}{|c|}{Schemes} & \multicolumn{2}{c|}{Single-Sleep} & \multicolumn{2}{c|}{Multiple-Sleep}\tabularnewline
\hline 
\hline 
\multicolumn{2}{|c|}{Metrics} & Simulation & Exact & Simulation & Exact\tabularnewline
\hline 
\multirow{3}{1cm}{PAoI} & Source 1 & 23.6114 & \textbf{23.5988} & 23.9390 & \textbf{23.9325}\tabularnewline
\cline{2-6} \cline{3-6} \cline{4-6} \cline{5-6} \cline{6-6} 
 & Source 2 & 17.8240 & \textbf{17.8309} & 18.0700 & \textbf{18.0696}\tabularnewline
\cline{2-6} \cline{3-6} \cline{4-6} \cline{5-6} \cline{6-6} 
 & Source 3 & 14.6042 & \textbf{14.6265} & 14.8052 & \textbf{14.8124}\tabularnewline
\hline 
\multirow{3}{1cm}{AoI} & Source 1 & 25.0400 & \textbf{25.0522} & 25.2873 & \textbf{25.2936}\tabularnewline
\cline{2-6} \cline{3-6} \cline{4-6} \cline{5-6} \cline{6-6} 
 & Source 2 & 18.9699 & \textbf{18.9782} & 19.1008 & \textbf{19.1058}\tabularnewline
\cline{2-6} \cline{3-6} \cline{4-6} \cline{5-6} \cline{6-6} 
 & Source 3 & 15.4433 & \textbf{15.4677} & 15.5284 & \textbf{15.5236}\tabularnewline
\hline 
\multicolumn{2}{|c|}{Energy Consumption Rate} & 1.8064 & \textbf{1.8065} & 1.8384 & \textbf{1.8384}\tabularnewline
\hline 
\end{tabular}\end{center}}

\caption{\label{tab:Simulation-Results-v.s.-1}Simulation results v.s. exact
results when using CS as the idling scheme. $\boldsymbol{\lambda}=(1.25,1.75,2.25)$
$H_{i}\sim Gamma(0.3,\frac{1}{0.3^{2}}),$ $\boldsymbol{\theta}=(0.3,0.4,0.5)$,
$U\sim Gamma(\frac{5}{4},1)$, $W\sim exp(0.5)$, $P_{B}=2.1$, $P_{ID}=1.1$,
$P_{SL}=0.3$, $P_{ST}=2.5$, $P_{DT}=0.6$. Simulation is performed
on sample paths of $3\times10^{6}$ regenerative cycles.}
\end{table*}

From Equations (\ref{eq:10-1}) and (\ref{eq:10-2}), we find that
the energy consumption rate depends on the idling scheme only through
the sleep probability $\theta_{i}^{CS}$. This phenomenon implies
that when replacing the idling scheme with BS and HT, we will have
the same expressions for the energy consumption rate. Moreover, from
Equations (\ref{eq:3}) and (\ref{eq:5}), we have 
\begin{eqnarray*}
\boldsymbol{E}[A_{i}] & = & \boldsymbol{E}[G_{i}]+\sum_{l=1}^{k}\frac{\lambda_{l}}{\lambda_{i}}\boldsymbol{E}[V_{l}]+\boldsymbol{E}[H_{i}].
\end{eqnarray*}
Since both $\boldsymbol{E}[G_{i}]$ and $\boldsymbol{E}[V_{i}]$ are
determined by the idling strategies through the sleeping probability,
we can conclude that when fixing the wakeup scheme (as one of the
N-policy, single-sleep scheme, or multiple-sleep scheme), using HT,
BS, and CS as the idling scheme will achieve the same PAoI. 

\section{Proof of Corollary \ref{cor:When-,-then}\label{sec:Proof-of-Corollary}}
\begin{IEEEproof}
For notation simplicity, we remove the superscript and let $\theta_{i}$
denote the sleeping probability under CS. Since both the denominator
and numerator of $\boldsymbol{E}[P^{CS}]$ are linear functions of
$N$, $\boldsymbol{E}[P^{CS}]$ is either increasing or decreasing.
We only need to compare the $\boldsymbol{E}[P^{CS}]$ in the cases
of $N=\infty$ and $N=1$. When $N=\infty$ and $\min_{i}\{\theta_{i}\}>0$,
we have $\boldsymbol{E}[P]=P_{SL}$. When $N=1$,  we have
\begin{eqnarray*}
 &  & \boldsymbol{E}[P^{CS}]\\
 & \geq & \frac{\sum_{i=1}^{k}\frac{\lambda_{i}}{\lambda}P_{SL}\bigg[\boldsymbol{E}[H_{i}]+\frac{1-\theta_{i}}{\lambda}+\theta_{i}(\frac{1}{\lambda}+\boldsymbol{E}[U])\bigg]}{\sum_{i=1}^{k}\frac{\lambda_{i}}{\lambda}\bigg[\boldsymbol{E}[H_{i}]+\theta_{i}\boldsymbol{E}[U]+\frac{1}{\lambda}\bigg]}\\
 & > & P_{SL}.
\end{eqnarray*}

Similarly, the minimal $\boldsymbol{E}[P]$ is achieved at either
$\theta_{i}=1$ or $\theta_{i}=0$ for all $i$ because the denominator
and numerator of $\boldsymbol{E}[P^{CS}]$ are linear on $\sum_{i=1}^{k}\lambda_{i}\theta_{i}$.
So the minimal $\boldsymbol{E}[P^{CS}]$ is achieved at either $\theta_{i}=0$
or $\theta_{i}=1$ for all $i$.

If $P_{ST}\leq P_{ID}$, then the minimal $\boldsymbol{E}[P^{CS}]$
is achieved at $\theta_{i}=1$ for all $i=1,...,k$. The reason is
that
\begin{eqnarray*}
\boldsymbol{E}[P^{CS}] & \geq & \frac{\sum_{i=1}^{k}\frac{\lambda_{i}}{\lambda}\bigg[P_{B}\boldsymbol{E}[H_{i}]+\frac{N}{\lambda}P_{SL}+\boldsymbol{E}[U]P_{ST}\bigg]}{\sum_{i=1}^{k}\frac{\lambda_{i}}{\lambda}\bigg[\boldsymbol{E}[H_{i}]+\frac{1}{\lambda}+\theta_{i}(\frac{N-1}{\lambda}+\boldsymbol{E}[U])\bigg]}\\
 & \geq & \frac{\sum_{i=1}^{k}\frac{\lambda_{i}}{\lambda}\bigg[P_{B}\boldsymbol{E}[H_{i}]+\frac{N}{\lambda}P_{SL}+\boldsymbol{E}[U]P_{ST}\bigg]}{\sum_{i=1}^{k}\frac{\lambda_{i}}{\lambda}\bigg[\boldsymbol{E}[H_{i}]+\boldsymbol{E}[U]+\frac{N}{\lambda}\bigg]}.
\end{eqnarray*}
The equality holds only when $\theta_{i}=1$ for all $i$. Hence proved. 
\end{IEEEproof}

\section{\label{sec:Proof-of-Corollary-1}Proof of Corollary \ref{cor:For-the-single}}
\begin{IEEEproof}
We can rewrite AoI under CS as 

\begin{eqnarray}
\boldsymbol{E}[\Delta^{CS}]=\frac{\eta+\beta N+\frac{\theta}{\lambda^{2}}N^{2}}{2(\gamma+\frac{\theta}{\lambda}N)}+\boldsymbol{E}[H]+\frac{\theta(1-U^{*}(\lambda))}{\lambda},
\end{eqnarray}
 with 
\begin{eqnarray}
\eta & = & H^{*(2)}(0)+2\bigg[\boldsymbol{E}[H]+H^{*(1)}(b)\bigg]\frac{1}{\lambda}+(1-\theta)\frac{2}{\lambda^{2}}\nonumber\\
& & -2H^{*(1)}(b)\boldsymbol{E}[U]+\theta\boldsymbol{E}[U^{2}],
\end{eqnarray}
\begin{eqnarray}
\beta & = & -2H^{*(1)}(b)\frac{1}{\lambda}+2\frac{\theta}{\lambda}\boldsymbol{E}[U]+\frac{\theta}{\lambda^{2}},
\end{eqnarray}
and

\begin{eqnarray}
\gamma & = & \boldsymbol{E}[H]+\frac{1}{\lambda}(1-\theta)+\theta\boldsymbol{E}[U].
\end{eqnarray}
We then have 
\begin{eqnarray}
\frac{\partial\boldsymbol{E}[\Delta^{CS}]}{\partial N} & = & \frac{\beta\gamma+2\frac{\theta}{\lambda^{2}}N\gamma+\frac{\theta^{2}}{\lambda^{3}}N^{2}-\eta\frac{\theta}{\lambda}}{2(\gamma+\frac{\theta}{\lambda}N)^{2}}.
\end{eqnarray}
The solution to $\frac{\partial\boldsymbol{E}[\Delta^{CS}]}{\partial N}=0$
is given by $N^{*}=2\frac{\theta}{\lambda^{2}}\gamma\sqrt{1-\frac{\lambda}{\gamma^{2}}(\beta\gamma-\eta\frac{\theta}{\lambda})}-2\frac{\theta}{\lambda^{2}}\gamma$.
From the facts that $\frac{\partial\boldsymbol{E}[\Delta]}{\partial N}|_{N=0}\leq0$
and $\frac{\partial\boldsymbol{E}[\Delta]}{\partial N}|_{N=\infty}>0$,
one can conclude that $N^{*}$ is the minimizer for $\boldsymbol{E}[\Delta^{CS}]$
if $N^{*}>0$. As $N$ increases from $0$ to $\infty$, $\boldsymbol{E}[\Delta^{CS}]$
decreases when $N\leq N^{*}$ and increases when $N>N^{*}$. This
implies that increasing the sleeping length (i.e., $N$) does not
always increase AoI. Note that $N^{*}$ is not always positive. For
$N^{*}$ to be negative, we need $\beta\gamma-\eta\frac{\theta}{\lambda}\geq0$,
which means 
\begin{align}
 & 2\boldsymbol{E}[He^{-bH}](\boldsymbol{E}[H]+\frac{1}{\lambda})+2\theta\boldsymbol{E}[U]\boldsymbol{E}[H]\nonumber\\
 &+(2-\theta)\frac{\theta}{\lambda}\boldsymbol{E}[U]+2\theta^{2}\{\boldsymbol{E}[U]\}^{2}\nonumber \\
\leq & \theta H^{*(2)}(0)+\frac{\theta\boldsymbol{E}[H]}{\lambda}+\frac{\theta(1-\theta)}{\lambda^{2}}+\theta^{2}\boldsymbol{E}[U^{2}].\label{eq:10}
\end{align}
Inequality (\ref{eq:10}) thus provides a sufficient condition under
which the AoI is not a monotone function of $N$. When $b=0$ for
CS, then Inequality (\ref{eq:10}) becomes 
\begin{eqnarray}
 &  & 2\{\boldsymbol{E}[H]\}^{2}+\frac{\boldsymbol{E}[H]}{\lambda}+2\boldsymbol{E}[U]\boldsymbol{E}[H]+\frac{\boldsymbol{E}[U]}{\lambda}+2\{\boldsymbol{E}[U]\}^{2}\nonumber\\
 & &\leq \boldsymbol{E}[H^{2}]+\boldsymbol{E}[U^{2}].\label{eq:10.1}
\end{eqnarray}
By letting $D=0$ for HT and $\theta^{BS}=1$ for BS we can also have
Inequality (\ref{eq:10.1}). One can easily show that Inequality (\ref{eq:10.1})
does not hold if the coefficients of variation (CV) of $H$ and $U$
are both smaller than 1 (i.e., $\frac{\sqrt{Var[H]}}{\boldsymbol{E}[H]}<1$
and $\frac{\sqrt{Var[U]}}{\boldsymbol{E}[U]}<1$). In such a case,
$N^{*}$ is a negative number, so increasing $N$ can increase AoI.
Therefore, whether enlarging the sleeping period length would increase
AoI depends on the CV of service and setup time distributions. When
the CV of $H$ and $U$ are large, it is possible that Inequality
(\ref{eq:10.1}) will hold. Hence proved. 
\end{IEEEproof}

\section{Proof of Theorem \ref{thm:The-optimal-}\label{sec:Proof-for-Theorem}}
\begin{IEEEproof}
With the Lagrangian multiplier $\boldsymbol{\eta}=(\eta_{1},\eta_{2},\eta_{3},\eta_{4})$,
we write the Lagrangian function of Problem \textbf{P1-relaxed} as
\begin{eqnarray*}
L(\boldsymbol{\eta},N,\theta)  &=&  \boldsymbol{E}[P^{CS}(N,\theta)]+\eta_{1}(\boldsymbol{E}[A^{CS}(N,\theta)]-\tau)\nonumber\\
& & +\eta_{2}(\theta-1)-\eta_{3}\theta-\eta_{4}(N-1).
\end{eqnarray*}
By the KKT condition for Problem \textbf{P1-relaxed}, the optimal
solution must satisfy 
\begin{eqnarray*}
 &  & \frac{\partial\boldsymbol{E}[P^{CS}(N,\theta)]}{\partial N}+\eta_{1}\frac{\partial\boldsymbol{E}[A^{CS}(N,\theta)]}{\partial N}-\eta_{4}=0;\\
 &  & \eta_{1}(\boldsymbol{E}[A^{CS}(N,\theta)]-\tau)=0;\\
 &  & \eta_{i}\geq0\mbox{ for }i\in\{1,2,3,4\}.
\end{eqnarray*}
From Corollaries \ref{cor:-is-an} and \ref{cor:When-,-then} we have
that for fixed $\theta>0$, then $\frac{\partial\boldsymbol{E}[P^{CS}(N,\theta)]}{\partial N}<0$
and $\frac{\partial\boldsymbol{E}[A^{CS}(N,\theta)]}{\partial N}>0$.
So by the KKT conditions above, we must have $\eta_{1}>0$, which
means $\boldsymbol{E}[A^{CS}(N,\theta)]-\tau=0$ always holds for
the optimal solution. So that in the optimization problem, the constraint
$\boldsymbol{E}[A^{CS}]\leq\tau$ is tight for the optimal solution
when $\theta>0$. Therefore, we can have $\theta=\frac{\tau-\frac{1}{\lambda}-2\boldsymbol{E}[H]}{\frac{N}{\lambda}+\boldsymbol{E}[U]-\frac{U^{*}(\lambda)}{\lambda}}$,
and the optimization problem \textbf{P1-relaxed} can be rewritten
as 
\begin{eqnarray*}
&   \min_{N<\infty}  \bigg\{\big[\frac{N}{\lambda}+\boldsymbol{E}[U]-\frac{U^{*}(\lambda)}{\lambda}\big]\big[P_{B}\boldsymbol{E}[H]+\frac{1}{\lambda}P_{ID}\big]\\
 &   +\big[\tau-\frac{1}{\lambda}-2\boldsymbol{E}[H]\big]\big[\frac{N}{\lambda}P_{SL}+\boldsymbol{E}[U]P_{ST}-\frac{P_{ID}}{\lambda}\big]\bigg\}\\
 &  \bigg/\bigg\{\big[\frac{N}{\lambda}+\boldsymbol{E}[U]-\frac{U^{*}(\lambda)}{\lambda}\big]\big[\boldsymbol{E}[H]+\frac{1}{\lambda}\big]\\
 &   +\big[\tau-\frac{1}{\lambda}-2\boldsymbol{E}[H]\big]\big[\frac{(N-1)}{\lambda}+\boldsymbol{E}[U]\big]\bigg\}\\
 &  \mbox{s.t. } N\geq\max\big\{1,\lambda(\tau-\frac{1}{\lambda}-2\boldsymbol{E}[H]-\boldsymbol{E}[U]+\frac{U^{*}(\lambda)}{\lambda})\big\}.
\end{eqnarray*}

We can then find that $\boldsymbol{E}[P^{CS}]$ has both denominator
and numerator as linear functions of $N$. Therefore, $\boldsymbol{E}[P^{CS}]$
is either an increasing or decreasing function of $N$. If it is decreasing,
then it is optimal to let $N$ become very large while keep $\theta=\frac{\tau-\frac{1}{\lambda}-2\boldsymbol{E}[H]}{\frac{N}{\lambda}+\boldsymbol{E}[U]-\frac{U^{*}(\lambda)}{\lambda}}>0$
and we have Type 1 solution. If $\boldsymbol{E}[P^{CS}]$ is an increasing
function of $N$, then the minimum $\boldsymbol{E}[P^{CS}]$ is achieved
when $N$ reaches the lower bound, where we have either the Type 2
or Type 3 solution. 
\end{IEEEproof}

\section{\label{sec:Proof-of-Corollary-2}Proof of Corollary\ref{cor:When-the-setup}}
\begin{IEEEproof}
As Type 1 solution is an asymptotic solution, when $N\rightarrow\infty$,
the energy consumption rate converges to
\begin{align}
\boldsymbol{E}[P_{Type1}^{CS}] & \triangleq\frac{P_{B}\boldsymbol{E}[H]+\frac{1}{\lambda}P_{ID}+(\tau-\frac{1}{\lambda}-2\boldsymbol{E}[H])P_{SL}}{\tau-\boldsymbol{E}[H]},
\end{align}
which is an energy consumption rate unrelated to $P_{ST}$. This shows
as long as the sleeping period length is large enough and the sleeping
probability is small enough, the effect of $P_{ST}$ on the energy
consumption rate becomes negligible. We now prove the corollary by
considering the following two cases:

Case I: When $\tau\geq\frac{1}{\lambda}+2\boldsymbol{E}[H]$, one
can easily verify that Type 3 solution does not exist. The energy
consumption rate for Type 2 solution is given by
\begin{eqnarray}
\boldsymbol{E}[P_{Type2}^{CS}|U=0]\triangleq\frac{P_{B}\boldsymbol{E}[H]+(\tau-\frac{1}{\lambda}-2\boldsymbol{E}[H])P_{SL}}{\tau-\boldsymbol{E}[H]}.
\end{eqnarray}
We can easily verify that $\boldsymbol{E}[P_{Type2}^{CS}|U=0]\leq\boldsymbol{E}[P_{Type1}^{CS}]$,
which means Type 2 solution will occur. 

Case II: When $\tau<\frac{1}{\lambda}+2\boldsymbol{E}[H]$, Type 2
solution does not exist. We can derive the energy consumption rate
for Type 3 solution as 
\begin{align}
\boldsymbol{E}[P_{Type3}^{CS}|U=0] & \triangleq\bigg\{(\tau-\frac{1}{\lambda}-2\boldsymbol{E}[H])(P_{SL}-P_{ID})\nonumber\\
&+P_{B}\boldsymbol{E}[H]+\frac{1}{\lambda}P_{ID}\bigg\}\big/(\boldsymbol{E}[H]+\frac{1}{\lambda}).
\end{align}
One can also verify that $\boldsymbol{E}[P_{Type3}^{CS}|U=0]\leq\boldsymbol{E}[P_{Type1}^{CS}]$.
Hence when there is no setup time, Type 1 solution does not occur. 
\end{IEEEproof}

\section{\label{sec:Discussion-for-the}Discussion for the Distribution of
$B_{i}$}

\begin{figure}
\centering{\includegraphics[scale=0.35]{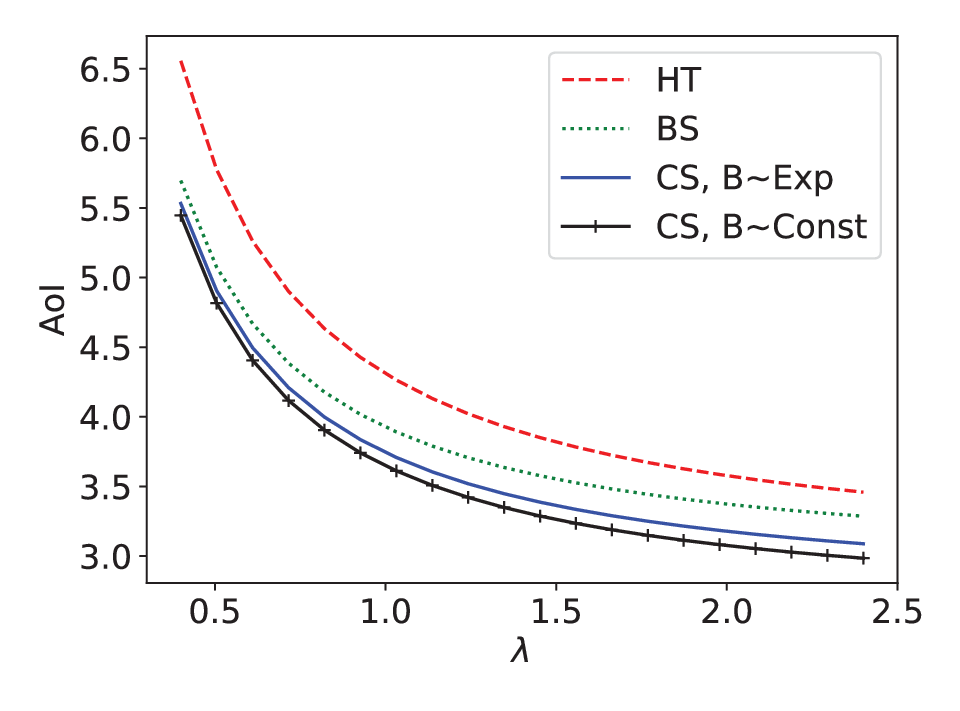}}
\caption{Idling scheme comparison with $N=2$ for N-policy, $k=1$, $H\sim Unif(0,\frac{5}{4}),$
$\theta=0.3$, $U\sim Gamma(\frac{5}{2},1)$, $D\sim constant$. \label{fig:Idling-Scheme-Comparison}}
\end{figure}

In our paper, we assume that the threshold variable $B_{i}$ is exponential
for CS. We make this assumption mainly to simplify our analysis, as
the terms $\theta_{i}^{CS}$ and $V_{i}^{*}(s)$ will have relatively
nice structures that can be written in closed form by the LST of $H_{i}$.
Setting $B_{i}$ as other distributions is also possible, but the
analysis and computation can be more complicated. When $B_{i}$ is
generally distributed with CDF $F_{B_{i}}(x)$, we have 
\begin{eqnarray}
\theta_{i}^{CS} =  \boldsymbol{P}(H_{i}<B_{i})=\int_{y=0}^{\infty}\int_{x=0}^{y}dF_{B_{i}}(y)dF_{H_{i}}(x)
\end{eqnarray}
and
\begin{align}
V_{i}^{*}(s) & = \int_{y=0}^{\infty}\int_{x=y}^{\infty}e^{-sx}\frac{\lambda}{s+\lambda}dF_{B_{i}}(y)dF_{H_{i}}(x)\nonumber\\
&+\int_{y=0}^{\infty}\int_{x=0}^{y}e^{-sx}(\frac{\lambda}{\lambda+s})^{N}U^{*}(s)dF_{B_{i}}(y)dF_{H_{i}}(x).
\end{align}
Specifically, if we set the threshold $b_{i}$ as a constant, then
\begin{eqnarray}
 \theta_{i}^{CS}=\boldsymbol{P}(H_{i}<b_{i})=F_{H_{i}}(b_{i})
\end{eqnarray}
and
\begin{eqnarray}
V_{i}^{*}(s) & = & \int_{x=b_{i}}^{\infty}e^{-sx}\frac{\lambda}{s+\lambda}dF_{H_{i}}(x)\nonumber\\
& & +\int_{x=0}^{b_{i}}e^{-sx}(\frac{\lambda}{\lambda+s})^{N}U^{*}(s)dF_{H_{i}}(x).
\end{eqnarray}
One needs to compute the numerical integration in order to obtain
$\theta_{i}^{CS}$ and $V_{i}^{*}(s)$. In Fig. \ref{fig:Idling-Scheme-Comparison},
we provide a comparison of idling schemes for the single source case.
We can observe that CS still has a smaller AoI than BS and HT when
the threshold is a constant. We also observe that when $B_{i}$ is
a constant, its AoI is a bit smaller than that when $B_{i}$ is exponential.
The reason is that by setting $B_{i}$ as a constant, the second moment
of $I_{ii}$ is reduced. However, we also note that the constant threshold
$B_{i}$ is not applicable in the scenario where the service time
$H_{i}$ is a constant.

\section{Discussion on the M/G/1/LCFS System\label{sec:Discussion-on-the}}
\begin{table*}[t!]
\centering{
\footnotesize{%
\begin{tabular}{|>{\raggedright}p{1.9cm}|c|c|c|c|c|c|c|c|c|c|}
\hline 
\multicolumn{2}{|l|}{$\lambda$} & 0.05 & 0.075 & 0.1 & 0.125 & 0.15 & 0.175 & 0.2 & 0.225 & 0.25\tabularnewline
\hline 
\hline 
\multirow{3}{1.9cm}{PAoI} & HT & 28.7801 & 21.7397 & 18.0365 & 15.7207 & 14.1763 & 13.0123 & 12.1018 & 11.4170 & 10.8217\tabularnewline
\cline{2-11} \cline{3-11} \cline{4-11} \cline{5-11} \cline{6-11} \cline{7-11} \cline{8-11} \cline{9-11} \cline{10-11} \cline{11-11} 
 & BS & 28.7958 & 21.7118 & 18.0436 & 15.7214 & 14.1768 & 13.0098 & 12.1041 & 11.3931 & 10.8325\tabularnewline
\cline{2-11} \cline{3-11} \cline{4-11} \cline{5-11} \cline{6-11} \cline{7-11} \cline{8-11} \cline{9-11} \cline{10-11} \cline{11-11} 
 & CS & 28.8098 & 21.7274 & 17.9993 & 15.7235 & 14.1647 & 12.9691 & 12.1045 & 11.4026 & 10.8307\tabularnewline
\hline 
\multirow{3}{1.9cm}{AoI} & HT & 19.6592 & 15.3973 & 13.3085 & 12.0023 & 11.1985 & 10.5868 & 10.0956 & 9.6929 & 9.3159\tabularnewline
\cline{2-11} \cline{3-11} \cline{4-11} \cline{5-11} \cline{6-11} \cline{7-11} \cline{8-11} \cline{9-11} \cline{10-11} \cline{11-11} 
 & BS & 11.2684 & 9.7029 & 9.0636 & 8.7560 & 8.6277 & 8.5135 & 8.4257 & 8.3603 & 8.2727\tabularnewline
\cline{2-11} \cline{3-11} \cline{4-11} \cline{5-11} \cline{6-11} \cline{7-11} \cline{8-11} \cline{9-11} \cline{10-11} \cline{11-11} 
 & CS & 10.5905 & 8.9904 & 8.3655 & 8.1135 & 8.0028 & 7.9372 & 7.9539 & 7.9405 & 7.9249\tabularnewline
\hline 
\multirow{3}{1.9cm}{Energy consumption rate} & HT & 1.3245 & 1.4899 & 1.6287 & 1.7377 & 1.8275 & 1.9018 & 1.9594 & 2.0067 & 2.0415\tabularnewline
\cline{2-11} \cline{3-11} \cline{4-11} \cline{5-11} \cline{6-11} \cline{7-11} \cline{8-11} \cline{9-11} \cline{10-11} \cline{11-11} 
 & BS & 1.3248 & 1.4915 & 1.6274 & 1.7384 & 1.8293 & 1.9016 & 1.9594 & 2.0049 & 2.0429\tabularnewline
\cline{2-11} \cline{3-11} \cline{4-11} \cline{5-11} \cline{6-11} \cline{7-11} \cline{8-11} \cline{9-11} \cline{10-11} \cline{11-11} 
 & CS & 1.3243 & 1.4914 & 1.6281 & 1.7388 & 1.8293 & 1.9023 & 1.9611 & 2.0067 & 2.0411\tabularnewline
\hline 
\end{tabular}
}}\caption{\label{tab:PAoI,-AoI,-and}PAoI, AoI, and energy consumption rate
comparison of HT, BS, and CS. Results are generated by simulation.
Each data point in the figure is generated by a simulated sample path
containing $10^{6}$ informative packets. $H\sim exp(0.4),$ $U=12.5$,
$N=1$, $\theta=0.3$.}
\end{table*}
\begin{figure}[h!]
\centering\includegraphics[scale=0.35]{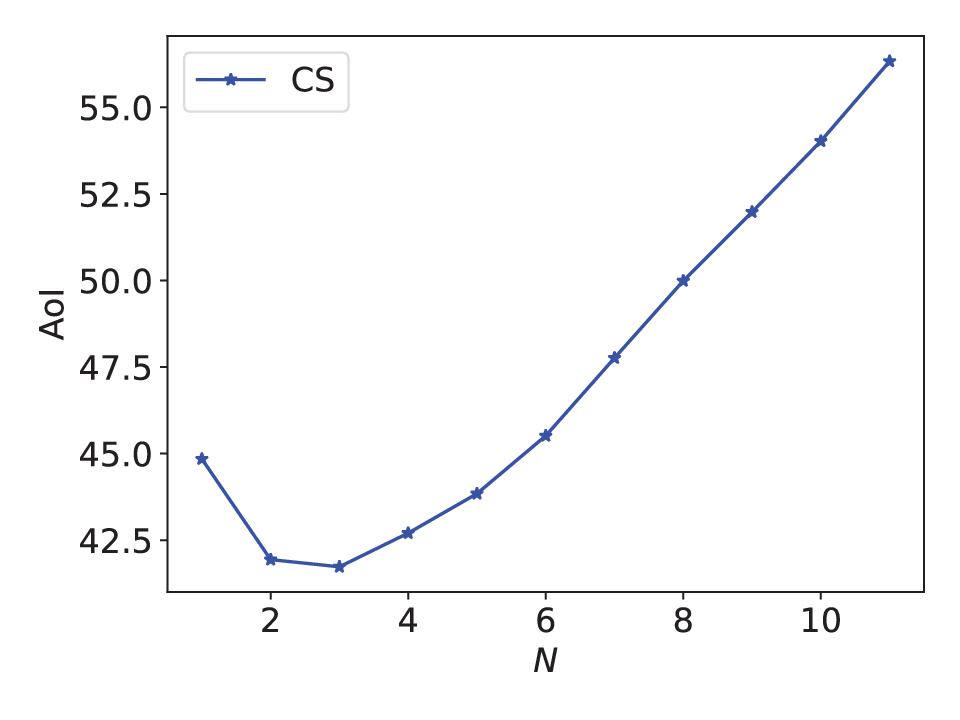}
\caption{\label{fig:AoI-in-M/G/1/LCFS}AoI in M/G/1/LCFS with CS and N-policy.
Results are generated by simulation. Each data point in the figure
is generated by a simulated sample path containing $10^{6}$ informative
packets. $\lambda=0.1$, $H\sim Gamma(\frac{1}{5},25)$, $U\sim Gamma(0.08,\frac{1}{0.08^{2}})$,
$\theta=1$.}
\end{figure}

{In the M/G/1/LCFS system, the server needs to decide
whether to sleep when the queue is empty. Under HT, the server will
idle for a period $D$ before the next arrival, so the probability
that the server sleeps after the queue becomes empty is the same as
that in the M/G/1/1 system, i.e., $\theta^{HT}=\boldsymbol{P}(D\leq L)=D^{*}(\lambda)$.
Under BS, the server sleeps with probability $\theta^{BS}$ when the
queue becomes empty, which is the same as that in the M/G/1/1 system.}

{Under HT, the sleeping probability in the M/G/1/LCFS
system will be slightly different from that in the M/G/1/1 system.
Specifically, the server will sleep if the length $H$ of the last
processed packet is smaller than the threshold $B$. For simplicity
of analysis, we still assume $B$ to be exponentially distributed
with rate $b$. The sleeping probability is thus $\theta=\boldsymbol{P}(H<B|H<L).$
To derive the sleeping probability, we first have 
\begin{align}
\boldsymbol{P}(H<x|H<L)  = & \frac{\int_{l=0}^{\infty}\boldsymbol{P}(H<x,H<l)\lambda e^{-\lambda l}dl}{\int_{l=0}^{\infty}\int_{h=0}^{l}dF_{H}(h)\lambda e^{-\lambda l}dl}\nonumber\\
  = & \frac{\int_{l=0}^{x}F_{H}(l)\lambda e^{-\lambda l}dl+F_{H}(x)e^{-\lambda x}}{H^{*}(\lambda)}.
\end{align}
Then the sleeping probability is 
\begin{eqnarray}
\theta & = & \int_{x=0}^{\infty}\frac{\int_{l=0}^{x}F_{H}(l)\lambda e^{-\lambda l}dl+F_{H}(x)e^{-\lambda x}}{H^{*}(\lambda)}be^{-bx}dx\nonumber\\
&=&\frac{H^{*}(b+\lambda)}{H^{*}(\lambda)}.
\end{eqnarray}
Based on the analysis in \cite{xu2021Information} and our derivation
for the M/G/1/1 system, we end up having the same closed-form expressions
of PAoI and energy consumption rate for HT, BS, and CS, which are
given as Equations (\ref{eq:13-1}) and (\ref{eq:13-2}).}

The closed-form expressions show that the PAoI and
energy consumption rate of the sleeping schemes are determined by
the sleeping probability $\theta$. However, these sleeping schemes
result in different AoI. A comparison of simulation results is provided
in Table \ref{tab:PAoI,-AoI,-and} of this response letter, from which
we see that once the sleeping probability is fixed, then HT, BS, and
CS have the same PAoI and energy consumption rate. The AoI under CS
is smaller than that under BS and CS. This observation is the same
as that for M/G/1/1 system. The reason is that both PAoI and energy
consumption rate are determined by first-order statistics $\theta$
and $N$, and AoI is determined by second-order statistics of the
idling and sleeping times. CS can reduce AoI since it avoids the variation
of idling and sleeping periods.
Since AoI is determined by second-order statistics
of the sleeping period in M/G/1/LCFS queue, extending the sleeping
period does not always increase AoI. In Fig. \ref{fig:AoI-in-M/G/1/LCFS},
we plot the AoI under CS and N-policy in the M/G/1/LCFS system, from
which we can see that AoI is not monotone on $N$. This observation
is the same as that for M/G/1/1 system. 

\end{appendices}
\end{document}